\documentclass[seceq,preprint]{ptptex}
\usepackage{wrapft}

\usepackage{graphicx}

\newcommand{\bra}[1]{{\!}\left\langle{}{ #1 }{}\right|{}}
\newcommand{\ket}[1]{{}\left|{}{ #1}{}\right\rangle{\!}}

\renewcommand{\vec}[1]{\mathbf #1}
\newcommand{\nn}{\nonumber}

\newcommand{\del}[2]{\frac{\partial #1}{\partial #2}}
\newcommand{\ddel}[3]{\frac{\partial^2 #1}{\partial #2 \partial #3}}
\newcommand{\sdel}[2]{\frac{\partial^2 #1}{\partial^2 #2}}

\newcommand{\sg}{\sigma^2}

\newcommand{\sgTf}{\bar{\sigma}^2_{t_{1i}}}
\newcommand{\sgTs}{\bar{\sigma}^2_{t_{2i}}}
\newcommand{\sgT}[1]{\bar{\sigma}^2_{T_D #1 }}
\newcommand{\dT}{{\Delta t^0_{1i}}}

\newcommand{\pC}{{\vec{p}_C}}

\newcommand{\pAz}{{\vec{p}_A^0}}
\newcommand{\pBz}{{\vec{p}_B^0}}
\newcommand{\pCz}{{\vec{p}_C^0}}
\newcommand{\pDz}{{\vec{p}_D^0}}

\newcommand{\xA}{\vec{X}_A}
\newcommand{\xB}{\vec{X}_B}
\newcommand{\xC}{\vec{X}_C}
\newcommand{\xD}{\vec{X}_D}
\newcommand{\tC}{T_C}
\newcommand{\tD}{T_D}
\newcommand{\vA}{\vec{v}_A}
\newcommand{\vB}{\vec{v}_B}
\newcommand{\vC}{\vec{v}_C}
\newcommand{\vD}{\vec{v}_D}

\title{Neutrino Oscillations in Intermediate States~II}
\subtitle{Wave Packets}

\author{Akinori Asahara$^1$, Kenzo Ishikawa$^1$, Takashi Shimomura$^1$\\ and Tetsuo Yabuki$^2$}

\inst{(1) Department of Physics, Faculty of Science, Hokkaido University Sapporo
060-0810, Japan \\
(2) Rakuno Gakuen University, Ebetsu 069-0836,Japan }

\abst{We analyze oscillations of intermediate neutrinos in terms of the scattering of particles
described by Gaussian wave packets. We study a scalar model as in a previous paper (I) but in
realistic situations, where the two particles of the initial state and final state are wave packets and
neutrinos are in the intermediate state. The oscillation of the intermediate neutrino is found from
the time evolution of the total transition probability between the initial state and final
state. The effect of a finite lifetime and a finite relaxation time are also studied. We find that
the oscillation pattern depends on the magnitude of wave packet sizes of particles in the initial
state and final state and the lifetime of the initial particle. For $\Delta m^2_{21}=3\times
10^{-2}$ eV$^2$, the oscillation probability deviates from that of the standard formula if the wave
packet sizes are around $10^{-13}$ m for $0.4$ MeV neutrino.}

\notypesetlogo

\begin{document}
\maketitle

\section{Introduction}
Neutrino oscillation is the only phenomenon in which to see effects of neutrino
masses at present. To analyze neutrino oscillations,  single particle wave functions have mainly
been studied. Because neutrino masses are very important, it is necessary to understand the quantum 
mechanics of neutrino oscillations in full detail. It is the purpose of the present paper to study
quantum mechanical aspects of neutrino oscillations beyond the single particle picture. We studied
particle oscillations from a nonstandard viewpoint in the framework of quantum field theory, where
neutrinos are in the intermediate state and the finite time interval effect is explicitly taken into
account, on the basis of plane waves in a previous paper.\cite{rf:1} \ We found that non-standard
oscillation patterns emerge in the exact plane waves. In many real physical processes, however,
particles are not exact plane waves but have finite  spatial extensions. Wave packets are suitable
to express these particles. We study particle oscillations in terms of the scattering amplitude of
the particles described by Gaussian wave packets. Neutrinos are in the intermediate state in this
amplitude, and neutrino oscillation is studied from this amplitude.

In scattering processes, the roles and importance of wave packets have been stressed by Goldberger
and Watson.~\cite{rf:15} The size of the wave packet of the initial state is determined from the
beam size, and it is a semi-micro scale for hadron beams. This corresponds to an energy
scale of the order of eV or less. By contrast, typical energy scales of hadron systems are of the
order of a few hundred MeV. Hence the effects of wave packets in the initial state are negligible in
standard hadron experiments. The size of the wave packet of the final state, on the other hand, is
determined by a detector. A detector is composed of many systems of materials. We regard the minimun set 
of materials for which a classical signal is taken as a unit detector. A unit detector is composed
of atoms and generates radiation, electrons, or other particles by which information from the quantum wave
function is transmitted to classical observers. The wave packet size is determined from the unit
detector and the resolution of the measurement is also determined from the unit detector. Hence the
wave packet size should be about the same as the spatial resolution of the detector. A unique value of
this size is not known now, so we study the dependence of the transition probability on the wave
packet sizes. The time resolution is determined from the time evolution of the detector. In this
paper, for simplicity, it is assumed that the time resolution is zero. Two observations at different
times are assumed to be independent. 

In neutrino oscillation experiments, the typical energy scale is extremely small and  the  spatial
sizes of experiments are of the order several hundreds km or more and are very different from scales
of ordinary experiments. Hence, the roles of wave packets in neutrino experiments are different from
their roles in other ordinary experiments and they should be clarified. It is one of our purposes in
the present work to study these problems. In fact, several theoretical works have been done on
Gaussian wave packets. But they are not sufficient. Especially, qualitative analysis is lacking. We
study effects that have not been studied to this time. The works in Refs.~\citen{rf:2,rf:3,rf:4}\
treat neutrinos as wave packets in a single particle picture. Other works treat the particles in the
initial and final states as wave packets in a field theoretical treatment.\cite{rf:5,rf:6}\ In the
latter, the standard S matrix theory, in which the transition time interval is set to infinity from
the beginning, was used, and the dependence of the amplitudes on neutrino parameters was
obtained. However, this standard treatment of the S matrix is inadequate in a process in which the
finite time interval effect is important and the dependence of amplitudes on external particles'
parameters are studied. The finite time interval effect becomes relevant in the situation in which the
intermediate particle is very light and interacts weakly with matter. Especially when the
intermediate particle consists of a superposition of several mass eigenstates and the mass squared
difference is very small, the finite time interval effect is not negligible. In the standard S
matrix theory, because the energy is strictly conserved and the interference of the amplitudes of
different mass does not occur, there is no oscillation when all the particles are exact plane
waves. In (I), the finite time interval effects in particle oscillations was shown to be important
when the observed particles are exact plane waves. Thus the finite time interval effect should be
important generally in the field theoretical treatment of neutrino oscillations. A modified S matrix
approach that allows us to investigate finite time interval effects should be applied for the study
of intermediate neutrino oscillations.

In the present paper, we extend our study of particle oscillations in the intermediate state to the
wave packet formalism. Oscillation amplitudes of neutrinos in the intermediate state where the
particles in the initial state and final state are described by wave packets are studied and effects
due to the finite wave packet, the finite time interval, and the finite lifetime or finite relaxation
time are found. Amplitudes are shown to deviate from the standard formula in extreme conditions,
when the wave packet sizes are very small. 

This paper is organized in the following manner. In Section 2 we give the general consideration of
the wave packet formalism in which the particles in the initial and final states are described by
Gaussian wave packets in the finite time interval method. In Section 3, the amplitude is computed
using the Gaussian approximation. In Section 4, we include a finite lifetime and a finite relaxation
time. In Section 5, numerical results in one spatial dimension are presented. A summary is given in
Section 6.

\section{The wave packet formalism}

Here we investigate neutrino oscillations in a scalar model in which particles in the initial and
final states have finite spatial widths that are described by Gaussian wave packets. Particles $A$,
$B$, $C$ and $D$ are external particles and are expressed by the field operators $\Phi_A(x)$,
$\Phi_B(x)$, $\Phi_C(x)$ and $\Phi_D(x)$. The fields $\Phi_{I_1}(x)$
and $\Phi_{I_2}(x)$ are mass eigenstates and are internal particles. The
Lagrangian density is given by
\begin{align}
L=\sum_{L=A,B,C,D} \left(\frac{1}{2}(\partial_\mu \Phi_L)^2-\frac{1}{2}{m_L}^2{\Phi_L}^2\right)
 +\sum_{i=1,2} \left(\frac{1}{2}(\partial_\mu \Phi_{I_i})^2-\frac{1}{2}{m_{I_i}}^2
 {\Phi_{I_i}}^2\right)
 -H_{\text{int}},
\end{align} 
where $m_{I_i}$ is the mass of $\Phi_{I_i}$ and $m_L$ that of $\Phi_L$. 
The interaction Hamiltonian is written 
\begin{align}
H_{\text{int}}&=H_{\text{int}}^1+H_{\text{int}}^2,\nonumber \\
H_{\text{int}}^1&=F_1 \int\!\!d^3 x\ \Phi_A(x) \Phi_{I_C}(x) \Phi_C(x),\nonumber \\
H_{\text{int}}^2&=F_2 \int\!\!d^3 x\ \Phi_B(x) \Phi_{I_D}(x) \Phi_D(x),
\end{align}
where $F_1$ and $F_2$ are coupling constants. The fields $\Phi_{I_C}(x)$ and $ \Phi_{I_D}(x)$ in the
above interaction Hamiltonian are linear combinations of the mass eigenstates :
\begin{align}
\Phi_{I_C}(x)&=\cos\theta \cdot \Phi_{I_1}(x)+\sin\theta \cdot
 \Phi_{I_2}(x), \nn \\
\Phi_{I_D}(x)&=-\sin\theta \cdot \Phi_{I_1}(x)+\cos\theta \cdot \Phi_{I_2}(x).
\end{align}
Here $\theta$ is the mixing angle between $\Phi_{I_C}$ and $\Phi_{I_D}$ and between 
$\Phi_{I_1}$ and $\Phi_{I_2}$. 

Each field operator is expanded in the interaction representation as 
\begin{align}
\Phi_L(x)&= \int \frac{d^3 p}{2E(p)_L} \exp\!\bigg( i\vec{p}\cdot\vec{x}-iE_L(p)t \bigg) \cdot a(\vec{p})_L +h.c.,\nn \\
E(\vec{p})_L&=\sqrt{\vec{p}^2 +{m_L}^2},
\end{align}
where $L$ stands for $A,B,C,D$ and $I_i$. 

We investigate the situation in which the particles $A$ and $B$ are prepared at time $t=0$ and
position $\vec{x}=\xA$, and the particles $C$ and $D$ are detected at $t=T_C$, $\vec{x}=\xC$ and
$t=T_D$, $\vec{x}=\xD$, respectively. (Fig.~\ref{diagram}) We assume $T_D > T_C$.
\begin{figure}[t]
 \begin{center}
  \scalebox{0.9}[1]{
\unitlength 0.1in
\begin{picture}( 57.2000, 17.9000)(  6.3000,-25.7500)
%
\special{pn 13}%
\special{pa 2012 1206}%
\special{pa 2226 1256}%
\special{fp}%
\special{sh 1}%
\special{pa 2226 1256}%
\special{pa 2166 1222}%
\special{pa 2174 1244}%
\special{pa 2158 1260}%
\special{pa 2226 1256}%
\special{fp}%
\put(38.6000,-24.0000){\makebox(0,0){{\large $(t_2,\vec{x}_2)$}}}%
\put(28.9000,-11.1000){\makebox(0,0){{\large $(t_1,\vec{x}_1)$}}}%
\put(54.5000,-23.8000){\makebox(0,0){{\large $(\tD,\xD)$}}}%
\put(22.8000,-24.9000){\makebox(0,0){{\large $(0,\xB)$}}}%
\put(45.1000,-9.8000){\makebox(0,0){{\large $(\tC,\xC)$}}}%
\put(14.4000,-8.7000){\makebox(0,0){{\large $(0,\xA)$}}}%
%
\special{pn 13}%
\special{pa 2760 1430}%
\special{pa 3700 2160}%
\special{da 0.070}%
%
\special{pn 13}%
\special{pa 1530 1070}%
\special{pa 2752 1416}%
\special{fp}%
%
\special{pn 13}%
\special{pa 3210 1270}%
\special{pa 3498 1184}%
\special{fp}%
\special{sh 1}%
\special{pa 3498 1184}%
\special{pa 3428 1184}%
\special{pa 3446 1200}%
\special{pa 3440 1222}%
\special{pa 3498 1184}%
\special{fp}%
%
\special{pn 13}%
\special{pa 2756 1424}%
\special{pa 3962 1028}%
\special{fp}%
%
\special{pn 13}%
\special{pa 2866 2266}%
\special{pa 3134 2232}%
\special{fp}%
\special{sh 1}%
\special{pa 3134 2232}%
\special{pa 3066 2222}%
\special{pa 3082 2240}%
\special{pa 3070 2260}%
\special{pa 3134 2232}%
\special{fp}%
%
\special{pn 13}%
\special{pa 2350 2332}%
\special{pa 3710 2160}%
\special{fp}%
%
\special{pn 13}%
\special{pa 4206 2228}%
\special{pa 4442 2270}%
\special{fp}%
\special{sh 1}%
\special{pa 4442 2270}%
\special{pa 4380 2238}%
\special{pa 4390 2260}%
\special{pa 4374 2278}%
\special{pa 4442 2270}%
\special{fp}%
%
\special{pn 13}%
\special{pa 3720 2160}%
\special{pa 4908 2334}%
\special{fp}%
\put(35.2000,-17.1000){\makebox(0,0){{\large $I$}}}%
\end{picture}%
}
 \end{center}
\caption{Diagram}
\label{diagram}
\end{figure}
The transition amplitude of the finite time interval $\tD$ between the initial 
and final states at second order is calculated as  
\begin{align}
\bra{\text{final}} S[t=\tD,t=0] \ket{\text{initial}}
= \bra{\text{final}} i^2 \int_{0}^{\tD}\!\!dt_2
 \int_{0}^{t_2}\!\!dt_1 H_{\text{int}} (t_1) H_{\text{int}} (t_2)
\ket{\text{initial}}. \label{tran_mat}
\end{align}

The initial state is composed of two particles, $A$ and $B$, which are the wave packets of the
finite spatial extents expressed by a distribution function. We assume that the initial time and the
final time are defined with infinite precision. This assumption makes the wave packet states defined
at different times independent of each other. The distribution function  of the momentum ${\vec p}$,
$ w({\vec p}_;\vec{p}^0,\vec{X},T; \sigma)$, has a finite extension around the central value
$\vec{p}^0$, and a Gaussian form of width $\sigma$ is assumed :
\begin{align}
w(\vec{p};\vec{p}^0,\vec{X},T;\sigma )
=\exp\bigg(-{\frac{(\vec{p} - \vec{p}^0 )^2}{{2\sigma}^2} } 
- i{\vec p} \cdot \vec{X}  + iE(\vec{p}) T\bigg).
\end{align}
The initial state is given by 
\begin{align}
\ket{\text{initial}}&= 
\int\!\!{{\frac{{d^{3}p_{A}} }{{\sqrt {(2\pi \sigma _{A}^{2} )^{3}}} }}} 
\int\!\!{{\frac{{d^{3}p_{B}} }{{\sqrt {(2\pi \sigma _{B}^{2} )^{3}}} }}}\
w({\vec p}_A;{\vec p}_{A}^0,\xA,0;\sigma _A )
\nonumber \\
&\times w({\vec p}_B;{\vec p}_{B}^0,\xB,0;\sigma _B )
a(\vec{p}_A)_A^\dagger  a(\vec{p}_B)_B^\dagger \ket{0}.\label{ini_state}
\end{align}
The final state is composed of two particles, $C$ and $D$, which are the wave packets. 
The final state is defined in the form
\begin{align}
\ket{\text{final}}&=
\int\!\!\frac{d^3 p_C}{\sqrt{(2\pi\sg_C)^3}}
\int\!\!{{\frac{{d^{3}p_{D}} }{{\sqrt {(2\pi \sigma _{D}^{2} )^{3}}} }}} 
 w({\vec p}_C;{\vec p}_C^0,{\vec X}_{C},T_C;\sigma _C )\nn\\
&\times w({\vec p}_D;{\vec p}_{D}^0,{\vec X}_{D},T_D;\sigma _D )
 a(\vec{p}_C)_C^\dagger a(\vec{p}_D)_D^\dagger \ket{0}.\label{fin_state}
\end{align}

In the above equations, $\xA$, $\xB$,$\xC$ and $\xD$ are the center positions of the wave packets,
and $\pAz$, $\pBz$, $\pCz$ and $\pDz$ are the central values of momentum of the particles $A$, $B$,
$C$ and $D$, respectively. We consider the spatial sizes of the wave packets, 
${\sigma_x}_L \equiv 1/(2\sigma_L)$~$(L=A,B,C,D)$, to be between a macroscopic size and a microscopic size. 

In the amplitude Eq.~\eqref{tran_mat}, the particles $A$, $B$, $C$ and $D$ represent directly
observed particles. The particle $I$ represent a scalar neutrino and appears only
in the intermediate state. Substituting $H_{\text{int}}(t)$, the amplitude
is given by
\begin{align}
S \equiv \bra{\text{final}}S[t=\tD,t=0]\ket{\text{initial}}
=\frac{1}{2}F_1 F_2 \sin 2\theta\left(S_1-S_2\right),
\end{align}
where
\begin{align}
S_i&=\left\{\prod_{L=A,B,C,D}
 \int\!\!\frac{d^3p_L}{\sqrt{(2\pi\sigma_L^2)^3}}
 \right\}
 \int\!\frac{d^3k}{(2\pi)^3 2E_i}
 \int_0^{\tD}\!\!dt_2\int_0^{t_2}\!\!dt_1 \int\!\!d^3x_2\int\!\!d^3x_1\nn\\
&\times e^{i(p_D-k_i-p_B)x_2+ i(p_C+k_i-p_A)x_1}
 w_A\ w_B\ w_C^\ast w_D^\ast. \label{amplitude}
\end{align}
From this amplitude, we study neutrino oscillation in the intermediate state. Because the neutrino
interacts with matter extremely weakly, it is not observed directly in real experiments. Hence, the
amplitudes of the present situation agree with the amplitudes of the realistic experimental situations.
  
Now we perform the integrations over $\vec{p}_A$, $\vec{p}_B$, $\vec{p}_C$ and $\vec{p}_D$ in
Eq.~\eqref{amplitude}. To integrate these variables, the energy is expanded around its central value,
\begin{align}
E_L({\vec p})&= E_L({\vec p}_L^0)+(\vec{p}-\vec{p}^0_L)\cdot\vec{v}_L ,\nn \\
\vec{v}_L&=\frac{\partial E_L(\vec{p}_L^0)}{\partial \vec{p}_L}, \label{def_velo}
\end{align}
where $\vec{v}_L$ is the velocity of the particle $L$ at this momentum, and the momentum integration 
is carried out as Gaussian integrations. After similar integrations over $\vec{x}_1$, $\vec{x}_2$ and $\vec{k}$, 
the transition amplitude is obtained as
\begin{align}
S_i&=N \exp\left[-\frac{\Delta\vec{P}^2}{2\sg}\right]
 \int^{\tD}_0\!\!dt_2\!\int^{t_2}_0\!\!dt_1
 \exp\bigg[-\frac{1}{2}Z_i(t_1,t_2,\tD)
 +it_1\Delta \tilde{E}_{1i}^0+it_2\Delta\tilde{E}_{2i}^0\bigg]\label{amp1},
\end{align}
where
\begin{align}
N&=\left(\frac{2\pi}{\sg}\right)^{3/2} e^{i\phi},\\
\phi&=-\xA\cdot\tilde{\vec{p}}_A^0-\xB\cdot\tilde{\vec{p}}^0_B
 +(\xC\cdot\tilde{\vec{p}}^0_C-\tC\tilde{E}_C^0)+(\xD\cdot\tilde{\vec{p}}^0_D-\tD\tilde{E}_D^0).
\end{align}
The quantity $Z_i$ in the exponent causes the amplitudes $S_i$ to have dominant contributions from
the regions $Z_i\simeq 0$ and is called a ``trajectory function''. The trajectory functions 
\begin{align}
Z_i(t_1,t_2,\tD)&=\frac{\sg_{AC}\sg_{BD}}{\sg}\vec{F}^2_i(t_1,t_2,\tD)
 +\frac{\sg_B\sg_D}{\sg_{BD}}\vec{G}^2(t_2,\tD)+\frac{\sg_A\sg_C}{\sg_{AC}}\vec{H}^2(t_1), \label{cls_trj}\\
\vec{F}_i(t_1,t_2,\tD)&=\vec{x}^0_2(t_2,T_D)-\vec{x}^0_1(t_1)-\vec{v}_i(t_2-t_1), \label{trj_f}\\
\vec{G}(t_2,\tD)&=\xD-\xB-t_2\vB-(\tD-t_2)\vD,\label{trj_g} \\
\vec{H}(t_1)&=\xC-\xA-t_1\vA-(\tC-t_1)\vC\label{trj_h}
\end{align}
give a classical particle picture, where $\vec{v}_i$ is the velocity of $I_i$ and is defined in the
same way as \eqref{def_velo}. Here we use for conciseness $\sg=\sg_A+\sg_B+\sg_C+\sg_D$ and 
$\sg_{ij}=\sg_i+\sg_j$ ($i,j=A,B,C,D$). The coefficient of $\vec{F}_i^2(t_1,t_2,\tD)$,
$\sg_{AC}\sg_{BD}/\sg$, is the width of the Gaussian function for the intermediate particle's
momentum. In our approach, the intermediate particles become wave packets automatically and then
sizes are given by those of the external particles.  

The momenta and energies appearing in $\phi$ are given by,
\begin{align}
\Delta\vec{P}&=\pCz+\pDz-\pAz-\pBz, \\
\tilde{\vec{p}}^0_L&=\vec{p}^0_L+\frac{\sg_L}{\sg}\Delta\vec{P}
 \quad\text{for}\quad L=A,B\\
\tilde{\vec{p}}^0_L&=\vec{p}^0_L-\frac{\sg_L}{\sg}\Delta\vec{P}
 \quad\text{for}\quad L=C,D,\\
\tilde{E}^0_L&=E^0_L+\frac{\sg_L}{\sg}\vec{v}_L\cdot\Delta\vec{P},
 \quad\text{for}\quad L=A,B\\
\tilde{E}^0_L&=E^0_L-\frac{\sg_L}{\sg}\vec{v}_L\cdot\Delta\vec{P},
 \quad\text{for}\quad L=C,D\\
\Delta\tilde{E}^0_{1i}&=\tilde{E}^0_C+E_i(\vec{k}^0)-\tilde{E}^0_A, \label{dif_e1}\\
\Delta\tilde{E}^0_{2i}&=\tilde{E}^0_D-E_i(\vec{k}^0)-\tilde{E}^0_B. \label{dif_e2}
\end{align}
The quantity $-\Delta \vec{P}^2/(2\sg)$ in the exponent of Eq.~\eqref{amp1} is called
the ``momentum function''. It gives a constraint on the differences between momenta.

The intermediate particles appear as wave packets even if they were not originally prepared as wave
packets, and their momenta are given by the following function of the momenta of external particles :
\begin{align}
\vec{k}^0&=\pDz-\pBz-\frac{\sg_{BD}}{\sg}\Delta\vec{P}.
\end{align}
$\vec{x}^0_1$ and $\vec{x}^0_2$ are the central positions of the interaction vertices and written
\begin{align}
\vec{x}^0_1(t_1)&=\frac{1}{\sg_{AC}}\bigg\{\sg_A(\xA+t_1\vA)+\sg_C(\xC-(\tC-t_1)\vC)\bigg\},\\
\vec{x}^0_2(t_2,T_D)&=\frac{1}{\sg_{BD}}\bigg\{\sg_B(\xB+t_2\vB)+\sg_D(\xD-(\tD-t_2)\vD)\bigg\}.
\end{align}
The Gaussian term in Eq.~\eqref{amp1} places the constraint $\pCz+\pDz-\pAz-\pBz \approx 0$ on the
momentum with the width $\sqrt{2\sg}$. The momentum is approximately conserved, because the initial
state and final state are approximate eigenstates of the momentum. The Gaussian terms in the integrand of 
$S_i$ place the constraints $\vec{F}_i\approx 0$, $\vec{G} \approx 0$ and $\vec{H} \approx 0$ on to
the times $t_1$ and $t_2$. These correspond to the classical trajectories in the particle picture. 
The latter constraints become stronger as the spatial sizes of the wave packets
become smaller.   

\section{Gaussian approximation in time integration}
In this section we study the amplitude Eq.~\eqref{amp1} further. In order to perform the time
integrations, the exponent of the integrand in Eq.~\eqref{amp1} is rewritten as follows
\begin{align}
&-\frac{1}{2}Z_i(t_1,t_2,\tD)+it_1\Delta \tilde{E}_{1i}^0+it_2\Delta\tilde{E}_{2i}^0 
 = -\frac{1}{2}Z_i(t^0_{1i},t^0_{2i},\tD)+it_{1i}^0(\tD)\Delta\tilde{E}_{1i}^0+it_{2i}^0(\tD)\Delta\tilde{E}_{2i}^0 \nn\\
&\quad -\frac{1}{2\sgTf}\left(t_1-t^0_{1i}(\tD)-\frac{\dT}{\bar{\sigma}_{t_{2i}}}(t_2-t^0_{2i}(\tD))
 -i\sgTf\Delta\tilde{E}^0_{1i}\right)^2\nn\\
&\quad -\frac{1}{2\sgTs}\left(t_2-t^0_{2i}(\tD) -i\bar{\sigma}_{t_{2i}}
 (\bar{\sigma}_{t_{2i}}\Delta\tilde{E}^0_{2i}+\dT\Delta \tilde{E}^0_{1i}) \right)^2\nn\\
&\quad -\frac{1}{2}\left(\bar{\sigma}_{t_{1i}}\Delta\tilde{E}^0_{1i}\right)^2
 -\frac{1}{2}\left(\bar{\sigma}_{t_{2i}}\Delta\tilde{E}^0_{2i}+\dT\Delta \tilde{E}^0_{1i}\right)^2,
\end{align}
where $\sgTf$ and $\sgTs$ are the widths of $t_1$ and $t_2$ and are given as
\begin{align}
\frac{1}{\sgTf} &=\frac{1}{2}\left(\frac{\partial^2 Z_i}{\partial t_1^2}\right),\label{sigt1}\\
\frac{1}{\sgTs} &=\frac{1}{2}\left(\frac{\partial^2 Z_i}{\partial t_2^2}\right)
 -\frac{1}{4}\sgTf\left(\frac{\partial^2 Z_i}{\partial t_1 \partial t_2}\right)^2,\label{sigt2}\\
\dT &= -\frac{1}{2}\sgTf\left
 (\frac{\partial^2 Z_i}{\partial t_1 \partial t_2}\right)\bar{\sigma}_{t_2}, \label{delt}
\end{align}
and $t^0_{1i}(T)$ and $t^0_{2i}(T)$ are regarded as the central values of the time of interactions
and are defined by
\begin{align}
\frac{\partial Z_i}{\partial t_1}\bigg|_{t_1=t^0_{1i} \atop t_2=t^0_{2i}} &= 0,\nn\\
\frac{\partial Z_i}{\partial t_2}\bigg|_{t_1=t^0_{1i} \atop
 t_2=t^0_{2i}} &= 0. \label{def_t0}
\end{align}
Because the explicit forms of $t^0_{1i}(\tD)$, $t^0_{2i}(\tD)$, $\sgTf$, $\sgTs$ and $\dT$ are quite
complicated, we just give quadratic forms of the trajectory function, and the explicit forms are
given in Appendix \ref{A}. Here the central values of the times and time widths depend on the
momenta of the external and internal particles. Therefore it happens in some situations that time
widths become very large, although the spatial widths of external particles are finite.

The amplitude $S_i$ is given by
\begin{align}
S_i&=N
 \exp\!\left[-\frac{1}{2}Z_i(t^0_{1i},t^0_{2i},\tD)
 -\frac{\Delta\vec{P}^2}{2\sg}
 -\frac{1}{2}\left(\bar{\sigma}_{t_{1i}}\Delta\tilde{E}^0_{1i}\right)^2
 -\frac{1}{2}\left(\bar{\sigma}_{t_{2i}}\Delta\tilde{E}^0_{2i}+\dT\Delta\tilde{E}^0_{1i}\right)^2
 \right] \nn\\
&\times \exp\!\left[it_{1i}^0(\tD)\Delta\tilde{E}_{1i}^0+it_{2i}^0(\tD)\Delta\tilde{E}_{2i}^0\right]\nn\\
&\times \int^{\tD}_0\!\!dt_2 \exp\!\left[-\frac{1}{2\sgTs}\left(t_2-t^0_{2i}(\tD) -i\bar{\sigma}_{t_{2i}}
 (\bar{\sigma}_{t_{2i}}\Delta\tilde{E}^0_{2i}+\dT\Delta\tilde{E}^0_{1i}) \right)^2\right]\nn\\
&\times \int^{t_2}_0 dt_1 \exp\!\left[ -\frac{1}{2\sgTf}\left(t_1-t^0_{1i}(\tD)-\frac{\dT}{\bar{\sigma}_{t_{2i}}}(t_2-t^0_{2i}(\tD))
 -i\sgTf\Delta\tilde{E}^0_{1i}\right)^2\right]. \label{amp2}
\end{align}
From the integrand in Eq.~\eqref{amp2}, we find that the integrations over $t_1$ and $t_2$ are 
separated when both time widths $\sgTf$ and $\sgTs$ are small enough compared to the time interval $T_D$. 
Afterwards, we assume that these conditions are satisfied. Then we can integrate Eq.~\eqref{amp2}
over $t_1$ and $t_2$, and we obtain
\begin{align}
S_i&=2\pi N \sqrt{\sgTf\sgTs}
 \times \exp\!\left[it_{1i}^0(\tD)\Delta\tilde{E}_{1i}^0+it_{2i}^0(\tD)\Delta\tilde{E}_{2i}^0\right]\nn\\
&\times \exp\!\left[-\frac{1}{2}Z_i(t^0_{1i},t^0_{2i},\tD)
 -\frac{\Delta\vec{P}^2}{2\sg}
 -\frac{1}{2}\left(\bar{\sigma}_{t_{1i}}\Delta\tilde{E}^0_{1i}\right)^2
 -\frac{1}{2}\left(\bar{\sigma}_{t_{2i}}\Delta\tilde{E}^0_{2i}+\dT\Delta\tilde{E}^0_{1i}\right)^2
 \right].\label{amp3}
\end{align}
The quantity $\sqrt{\sgTf\sgTs}$ appears as an overall factor in a consequence of the time
integrations. The amplitudes become large when the time widths are large. This factor is derived
only in a field theoretical treatment in which whole process is involved.

In the second line of Eq.~\eqref{amp3}, the contribution from the trajectory 
function to the amplitudes becomes maximal at $t^0_{1i}$ and $t^0_{2i}$.
Here, $t^0_{1i}$ is the central value of the production time of $I_i$, and $t^0_{2i}$ is
that of the detection time. The Gaussian integration with $t_{1i}^0$ and $t_{2i}^0$ in
Eq.~\eqref{amp2} becomes negligible, unless the condition
\begin{align}
0 < t_{1i}^0(\tD) < t_{2i}^0(\tD) < \tD. \label{eq2}
\end{align}
is satisfied.

We see from Eq.~\eqref{amp3} that the phase difference $\Theta_{21}$ up to
$\mathcal{O}(m^2)$ can be expressed in the form
\begin{align}
\Theta_{21}(T)=-\frac{\Delta m^2_{21}}{2|\vec{k}^0|}(t^0_2(\tD)-t^0_1(\tD))
 +\Delta m^2_{21}\left(\Delta\tilde{E}^0_1\del{t^0_{1i}}{m^2_i}\bigg|_{m_i=0}
 +\Delta\tilde{E}^0_2\del{t^0_{2i}}{m^2_i}\bigg|_{m_i=0}\right),\label{phase_dif}
\end{align}
where $t^0_1$, $t^0_2$, $\Delta \tilde{E}^0_1$ and $\Delta \tilde{E}^0_2$ are
the central times and energy differences with $m_i=0$.

The first term of Eq.~\eqref{phase_dif} corresponds to the phase of the standard formula, and the
second term results from the field theoretical treatment. The phase difference takes a form similar
to that of the standard formula when $\Delta \tilde{E}^0_1$, $\Delta \tilde{E}^0_2$ and 
$\Delta \vec{P}$ are zero. In this case, we have the standard formula when $t^0_2-t^0_1$ can be
regarded as the travel time of the neutrinos. However, $t^0_2$ is different from the final time 
$\tD$, and $t_1^0 $ is different from the 
initial time, and they are given by Eq.~\eqref{def_t0}. Thus, our formula is not exactly the same as
the standard formula. 

The absolute square of the amplitude, $|S|^2$, gives the transition probability from the initial state
\eqref{ini_state}, which is prepared at $t=0$ to the final state \eqref{fin_state}, which is measured  
at $t=\tD$. In observations of solar neutrinos and atmospheric neutrinos, the detection time is not
measured. In baseline neutrino experiments, the detection time is measured but with finite
precision. Therefore we have to sum up $|S|^2$ in a finite detection time. Here we make the
assumption that the measurements at different times are independent phenomena. This assumption is
consistent with the definition of wave packet states that they are defined with infinite precision 
and with the fact that the event rate is very small. From this assumption, we can integrate $|S|^2$ over 
the detection time $\tD$ :
\begin{align}
|\tilde{S}|^2 = \int\!d\tD\ |S|^2.
\end{align}
The integration interval is different for different situations. It is from 0 to $\infty$ for
neutrinos from the sun or atmosphere,  and from $\tD-\Delta \tD/2$ to $\tD+\Delta \tD/2$ for
baseline neutrino oscillation experiments, where $\Delta \tD$ is the resolution of the detection
time. 

We assume that the Gaussian approximation is valid for the $\tD$ integral. Then, the transition
probability becomes 
\begin{align}
|\tilde{S}|^2 =& 
\sum_{i=1,2}\sqrt{\sgT{i}}C_i^2 \exp[2A_i]-2C_1C_2\sqrt{\frac{2\sgT{1}\sgT{2}}{\sgT{1}+\sgT{2}}}\nn\\
&\hspace{1.0cm} \times \exp\left[A_1+A_2-\frac{({T^0_D}_2-{T^0_D}_1)^2}{2(\sgT{1}+\sgT{2})}
 -\frac{1}{2}\tilde{\sigma}^2_{T_D}\left(\del{\Theta_{21}(\tD)}{\tD}\right)^2\right]
 \cos(\Theta_{21}(\tilde{T}_D^0)),\label{prob}
\end{align}
where $A_i$ represents the energy and momentum functions and trajectory function, and $\sgT{i}$
represents the width of the detection time, $\tD$. These are given by 
\begin{align}
C_i&=(2\pi)^{5/2}\ (\sg)^{-3/2}\ (\sgTf\sgTs)^{1/2}\ \frac{1}{E_i}, \nn\\
A_i&=-\frac{1}{2}Z_i(t^0_{1i},t^0_{2i},{T^0_D}_i)-\frac{\Delta\vec{P}^2}{2\sg}
 -\frac{1}{2}\left(\bar{\sigma}_{t_{1i}}\Delta\tilde{E}^0_{1i}\right)^2
 -\frac{1}{2}\left(\bar{\sigma}_{t_{2i}}\Delta\tilde{E}^0_{2i}+\dT\Delta\tilde{E}^0_{1i}\right)^2,\label{exp_a} \\
\tilde{\sigma}_{T_D}&=\sqrt{\frac{\sgT{1}\sgT{2}}{\sgT{1}+\sgT{2}}},\\
\frac{1}{\sgT{i}}&=\frac{1}{2}\left(\frac{\partial^2 Z_i}{\partial {T_D}^2} \right). \label{sigTd}
\end{align} 
The quantity $\Theta_{21}$ is the phase difference between $I_1$ and $I_2$ in
Eq.~\eqref{phase_dif}. The detection time is replaced by its central
value,
\begin{align}
\tilde{T}_D^0=\frac{\sgT{1}{T^0_D}_2+\sgT{2}{T_D^0}_1}{\sgT{1}+\sgT{2}},\nn\\
\intertext{where ${T^0_D}_i$ is the central value of the detection time in which only the mass 
eigenstate $I_i$ appears as the intermediate state, and it is defined by}
\frac{\partial Z_i}{\partial T_D}\bigg|_{T_D={T^0_D}_i}=0.
\end{align}

In the exponent of the second term in Eq.~\eqref{prob}, last two terms are
characteristic terms in wave-packet treatment. One is called the ``decoherence function'',  
which represents an overlap in the detection time through two
intermediate states :
\begin{align}
\text{Decoherence function}= -\frac{({T^0_D}_2-{T^0_D}_1)^2}{2(\sgT{1}+\sgT{2})}.\label{decoherence}
\end{align}
When the detection time difference, ${T^0_D}_2-{T^0_D}_1$, becomes larger than
the detection time width, $\sqrt{\sgT{1}+\sgT{2}}$, the oscillation
disappears, because the coherence in the time direction is lost.
The other is called the ``phase function'', which gives the
condition that the detection time width must be smaller than the
oscillation period :
\begin{align}
\text{Phase function}
 &=-\frac{1}{2}\tilde{\sigma}^2_{T_D}\left(\del{\Theta_{21}(T_D)}{T_D}\right)^2\nn\\
 &=-\frac{1}{2}\left(\frac{\tilde{\sigma}_{T_D}}{{T_D}^{\text{osc}}}\right)^2\big(2\pi-\Theta_{21}(0)\big)^2.
 \label{phase-func}
\end{align}
Here $T_D^{osc}$ is a period and is defined by $\Theta_{21}(T_D^{osc})=2\pi$.
The extra coefficient $2\pi-\Delta\Theta(0)$ appears due to the field
theoretical treatment. 

The energy function places a constraint on the energy differences 
$\Delta \tilde{E}^0_{1i}$ and $\Delta \tilde{E}^0_{2i}$,
\begin{align}
-\frac{1}{2}\left(\bar{\sigma}_{t_{1i}}\Delta\tilde{E}^0_{1i}\right)^2
 -\frac{1}{2}\left(\bar{\sigma}_{t_{2i}}\Delta\tilde{E}^0_{2i}+\dT\Delta\tilde{E}^0_{1i}\right)^2.
\end{align}
Energy conservation is satisfied when the energy function is zero.

When the energy-momentum and the trajectory functions in one dimension are zero,
the phase difference, Eq.~\eqref{phase_dif}, becomes 
\begin{equation}
\Theta_{21}(\tilde{T}^0_D)=\frac{\Delta m^2_{21}}{2|\vec{k}^0|}
 (X_B-X_A+t^0_2(\tilde{T}^0_D)v_B-t^0_1(\tilde{T}^0_D)v_A).\label{phase_std}
\end{equation}
Equation \eqref{phase_std} agrees with the phase of the standard
formula\cite{rf:8,rf:9}\ if $t^0_2v_B$ and $t^0_1v_A$ are negligible.

\section{The effect of a finite lifetime on the wave packets}
In this section, we study the transition amplitude and probability when a 
source particle has a finite lifetime.~\footnote{This includes cases in which a particle is stable
but the state looses the quantum mechanical coherence after a finite time. The $\tau $ stands for
the relaxation time in this case.} \ From these, the particle oscillation of the intermediate
particle is studied. 

The finite lifetime $\tau$ is introduced for the particle A in the 
following manner. 
The field operator of the particle $A$, $\Phi_A$, contains $\Gamma=1/\tau$ as
\begin{align}
\Phi_A(x) = \int\!\frac{d^3p}{2 E_A}\ 
\exp\big(i\vec{p}\cdot\vec{x} -i(E_A-i\Gamma/2)t\big)\ 
a(\vec{p})_A+h.c. 
\end{align}
In consequence of this addition in Eq.~\eqref{amp1}, a damping factor $-\Gamma/2$ is 
added to $E_A(p^0_A)$ in the previous section. In this section, we study the amplitude and
probability in the case that both widths $\bar{\sigma}_{t_2}$ and $\bar{\sigma}_{t_1}$ are small
enough compared to the time interval $\tD$ and the lifetime $\tau$ satisfies $\tau \gg \bar{\sigma}_{t_1}$ 
and $\bar{\sigma}_{t_2}$. Then, as in the previous section, it has been found that the integrations
over $t_1$ and $t_2$ in the amplitudes are separated.  

If these conditions are not satisfied, the Gaussian approximation is invalid, and the integrations
over $t_1$ and $t_2$ cannot be separated. Then, the calculation would be similar to that for in
plane waves. Below, we assume that these conditions are satisfied. Therefore the transition
amplitude given by Eq.~\eqref{amp2} are modified into the following form:
\begin{align}
S&=\frac{1}{2}F_1F_2\sin 2\theta
 \big(S_1-S_2\big), \label{amp_life}
\end{align}
where
\begin{align}
S_i&=2\pi N\sqrt{\sgTf\sgTs}\times 
 \exp\!\left[i{t_{1i}^0}'(\tD)\Delta\tilde{E}_{1i}^0+i{t_{2i}^0}'(\tD)\Delta\tilde{E}_{2i}^0
 -{t_{1i}^0}'\frac{\Gamma}{2} \right] \nn\\
&\times \exp\!\left[-\frac{1}{2}Z_i({t^0_{1i}}',{t^0_{2i}}',\tD)
 -\frac{\Delta\vec{P}^2}{2\sg}
 -\frac{1}{2}\left(\bar{\sigma}_{t_{1i}}\Delta\tilde{E}^0_{1i}\right)^2
 -\frac{1}{2}\left(\bar{\sigma}_{t_{2i}}\Delta\tilde{E}^0_{2i}+\dT\Delta\tilde{E}^0_{1i}\right)^2
 \right].\label{amp4}
\end{align}
The new central times ${t_{1i}^0}'$ and ${t_{2i}^0}'$ of the Gaussian functions of $t_{1} $ and
$t_{2} $ in Eq.~\eqref{amp4} are obtained from old ones in Eq.~\eqref{def_t0} as follows
\begin{align}
{t_{1i}^0}'(\tD)&=t_{1i}^0(\tD)-(\sgTf+{\dT}^2)\frac{\Gamma}{2}, \\
{t_{2i}^0}'(\tD)&=t_{2i}^0(\tD)-\bar{\sigma}_{2i}\dT\frac{\Gamma}{2}.
\end{align}
The transition probability is also calculated by assuming the Gaussian
approximation for $\tD$ integration. It is given by
\begin{align}
|\tilde{S}|^2 =&\sum_{i=1,2}\sqrt{\sgT{i}}\tilde{S}_i^2
 -2\sqrt{\frac{2\sgT{1}\sgT{2}}{\sgT{1}+\sgT{2}}} \tilde{S}_1 \tilde{S}_2\nn\\
&\hspace{1cm}\times \exp\left[-\frac{({{T^0_D}_2}'-{{T^0_D}_1}')^2}{2(\sgT{1}+\sgT{2})}
 -\frac{1}{2}\tilde{\sigma}^2_{T_D}\left(\del{\Theta'_{21}(\tD)}{\tD}\right)^2\right]
 \cos(\Theta'_{21}(\tilde{T}_D^0)),\label{prob2}\\
\tilde{S}_i=&C_i \exp\left[A_i'-\frac{\Gamma}{2} {t^0_{1i}}'({\tD^0}_i)\right],\label{amp-si}
\end{align}
where the coefficients $C_i$ are the same as before. ${T_D^0}'_i$ is the central time of the
detection time and is given by  
\begin{align}
{\tD^0}'_i= {\tD^0}_i-\frac{\Gamma}{2}\sgT{i}\left(\del{{t_{1i}^0}'(\tD)}{\tD}\right).
\end{align}
$\Theta'_{21}$ and $A_i'$ are obtained by replacing
$t^0_{1i}$, $t^0_{2i}$ and ${T^0_D}_i$ by ${t^0_{1i}}'$, ${t^0_{2i}}'$ and
${T^0_D}'_i$ in Eq.~\eqref{phase_dif} and \eqref{exp_a}. The Gaussian approximation for 
$t_1$ and $t_2$ is useful and is valid when $\tau \gg \sgTf$, $\sgTs$ is satisfied. 
This condition is satisfied for a lifetime $\tau=10^{-8}$ sec of the pion and ${\sigma_x}_L$~
$(L=A,B,C,D)$ of atomic size. 

We see from Eqs.~\eqref{prob} and \eqref{prob2} that the exponential in the
oscillation term becomes maximum at the given positions $X_i$ when the energy-momentum function and
the trajectory function vanish. However the peak of the oscillation probability does not always
coincide with that of the exponential, because of the coefficient and the lifetime. The lifetime and
relaxation time reduce the magnitude of the oscillation probability depending on how long the
particle $A$ lives. If the constraints from energy-momentum conservation and the classical
trajectories are weak, the oscillation probabilities become maximal at the position where the energy
and momentum are not conserved. Then, as a result, the oscillation length changes from that of the
standard formula. Such situations seem quite strange. However as we show below, these phenomena occur 
when the spatial widths of the external particles are extremely small, on the order of $1$ fm.

\section{Coherence conditions} 
Here we examine necessary conditions for oscillation of an intermediate particles to take place,
based on the transition probabilities Eq.~\eqref{prob} and \eqref{prob2}. Oscillation occurs when
the interference term of two amplitudes becomes finite. Two  amplitudes have peaks in  different positions  and 
decrease rapidly, so the interference term becomes finite only  when the peak
position overlaps within the widths. We find these  conditions, coherent 
conditions, in the present section. There exist several previous works on this 
topic ,~\cite{rf:2}\tocite{rf:6}\cite{rf:12,rf:13}\ but our results based on field theory 
are different from them. In our method, the coherence conditions are written in terms of measured quantities, 
like the positions, velocities and wave packet sizes for external particles. 

\subsection{Energy function}
The factor
\begin{align}
\exp\left[-\frac{1}{2}\left(\bar{\sigma}_{t_{1i}}\Delta\tilde{E}^0_{1i}\right)^2
 -\frac{1}{2}\left(\bar{\sigma}_{t_{2i}}\Delta\tilde{E}^0_{2i}+\dT\Delta\tilde{E}^0_{1i}\right)^2\right],
 \label{e_cons}
\end{align}
which also appears in the amplitude $S_i$,
yields the approximate energy conservation and imposes 
the constraint on $\Delta m_{21}^2$ and $\sigma_x$ to 
generate oscillations. The Gaussian function (\ref{e_cons}) has a
width $\sgTf$ of $\Delta \tilde{E}_{1i}$ and $\sgTs$ of 
$\Delta \tilde{E}_{2i}+\frac{\dT}{\sgTs}\Delta \tilde{E}_{1i}$. Note that the second term in \eqref{e_cons} contains 
$\Delta \tilde{E}_{1i}$ because of finite interaction time interval. We understand from
these terms that when $\Delta \tilde{E}_{12}-\Delta \tilde{E}_{11}$ or $\Delta \tilde{E}_{22}-\Delta \tilde{E}_{21}$ 
becomes larger than $(\bar{\sigma}_{t_{1i}})^{-1}$ or
$(\bar{\sigma}_{t_{2i}}-\Delta t_{1i})^{-1}$, the interference of $I_1$ and
$I_2$ disappears, and the oscillation does not take place. From Eqs.~\eqref{dif_e1} and 
\eqref{dif_e2}, 
$\Delta \tilde{E}_{12}-\Delta \tilde{E}_{11}=\frac{\Delta m^2_{21}}{2|\vec{k}^0|}$ 
and $\Delta \tilde{E}_{22}-\Delta \tilde{E}_{21}=-\frac{\Delta m^2_{21}}{2|\vec{k}^0|}$, 
this coherence condition for oscillations is reduced to
\begin{align}
\frac{\Delta m_{21}^2}{2E} &\le \frac{1}{|\bar{\sigma}_{t_1}|},\nn\\
\frac{\Delta m_{21}^2}{2E} &\le \frac{1}{|\bar{\sigma}_{t_2}-\Delta t_1|},\label{coh_cnd1} 
\end{align}
where $E$ represents $|\vec{k}^0|$ and $\bar{\sigma}_{t_1}$, $\bar{\sigma}_{t_2}$ and 
$\Delta t_1$ are the values of Eq.~\eqref{sigt1}--\eqref{delt} at $m_i=0$. 
The above coherence conditions (\ref{coh_cnd1}) are expressed as
\begin{align}
\frac{2\pi}{L_{\text{osc}}} \le \frac{1}{|\bar{\sigma}_{t_1}|},
\quad \text{or} \quad \frac{1}{|\bar{\sigma}_{t_2}-\Delta t_1|},\label{eq3}
\end{align}
where $L_{\text{osc}} = \frac{4\pi E}{\Delta m^2_{21}}$ is the oscillation
length in the standard oscillation formula, which is almost the same as that derived from 
\eqref{phase_std}. Using the spatial widths $\bar{\sigma}_{x_1}$ and
$\bar{\sigma}_{x_2}$ of intermediate particles in the production and detection processes defined by 
\begin{align}
\bar{\sigma}_{x_1} &\equiv v_I |\bar{\sigma}_{t_1}|, \nn\\
\bar{\sigma}_{x_2} &\equiv v_I |\bar{\sigma}_{t_2}-\Delta t_1|,\label{av_sig_x}
\end{align}
where $v_I$ is the velocity of the intermediate particle and is equal to $1$, Eqs.~\eqref{eq3} is written as
\begin{align}
\bar{\sigma}_{x_1},\ \bar{\sigma}_{x_2} &\le \frac{L_{\text{osc}}}{2\pi} \label{const_av_sig_x}
\end{align}
for standard oscillations to take place. When the above constraints are not satisfied, 
ordinary interference of $I_1$ and $I_2$ does not exist, and the oscillation 
disappears. Consequently, the transition probability becomes constant in 
the time interval $\tD$.

\subsection{Trajectory function}
The trajectory functions in each amplitude, $S_i$,
\begin{align}
&\exp\left[-\frac{1}{2}Z_i({t^0_1}_i,{t^0_2}_i,{\tD^0}_i)\right] \nn \\
&=\exp\left[-\frac{\sg_{AC}\sg_{BD}}{2\sg}\vec{F}^2_i({t^0_1}_i,{t^0_2}_i,{\tD^0}_i)
 -\frac{\sg_B\sg_D}{2\sg_{BD}}\vec{G}^2({t^0_2}_i,{\tD^0}_i)-\frac{\sg_A\sg_C}{2\sg_{AC}}\vec{H}^2({t^0_1}_i)\right],
\end{align}
also give necessary constraints for oscillation to take place. These forms are very 
complicated, and it is difficult to find expressions using parameters of external particles. 
For this reason, we express constraints using the center times and the center positions. 
These conditions are given as
\begin{align}
|\delta \vec{F}|&\equiv |\vec{F}_2-\vec{F}_1| \le \sqrt{\frac{\sg}{\sg_{AC}\sg_{BD}}}\ , \\
|\delta \vec{G}|&\equiv |\vec{G}_2-\vec{G}_1| \le \sqrt{\frac{\sg_{BD}}{\sg_{B}\sg_{D}}}\ , \\
|\delta \vec{H}|&\equiv |\vec{H}_2-\vec{H}_1| \le \sqrt{\frac{\sg_{AC}}{\sg_{A}\sg_{C}}}\ .
\end{align}
From Eqs.~\eqref{trj_f} to \eqref{trj_h}, the above constraints are rewritten as
\begin{align}
\bigg|\delta \vec{x}^0_2 -\delta \vec{x}^0_1 -\hat{\vec{k}}(\delta t^0_2-\delta t^0_1)
+\frac{\Delta m^2_{21}}{2E^2}\hat{\vec{k}}\big(t^0_2({\tD^0}_i)-t^0_1({\tD^0}_i)\big)\bigg| 
&\le \sqrt{\frac{\sg}{\sg_{AC}\sg_{BD}}}\ , \label{traj-cnstr-f}\\
\big|(\vB-\vD) \delta t^0_2+\vD\delta T^0\big| &\le
 \sqrt{\frac{\sg_{BD}}{\sg_{B}\sg_{D}}}\ ,\label{traj-cnstr-g}\\
\big|\vC-\vA\big|\big|\delta t^0_1\big| &\le \sqrt{\frac{\sg_{AC}}{\sg_{A}\sg_{C}}}\ \label{traj-cnstr-h}.
\end{align}
Throughout this section, we write the center times without a prime for the case in which 
the particle $A$ has a finite lifetime. 

The right-hand side of the first constraint, Eq.~\eqref{traj-cnstr-f} is the spatial width 
of the intermediate particle, and those of the second and third constraints, Eq.~\eqref{traj-cnstr-g} and 
\eqref{traj-cnstr-h}, are the sums of the spatial widths of the external particles. 
On the left-hand side of the first condition, 
$\frac{\Delta m^2_{21}}{2E^2}\hat{\vec{k}}(t^0_2-t^0_1)$ is rewritten as follows
\begin{align}
\lambda\frac{L_{\text{travel}}}{L_{\text{osc}}}\hat{\vec{k}}, \label{delta-f}
\end{align}
where $L_{\text{travel}}=t^0_2-t^0_1$, and  
$\lambda$ is the de Broglie wavelength for the intermediate particles. 
Unless $L_{\text{travel}}$ is much larger than $L_{\text{osc}}$, Eq.~\eqref{delta-f} is 
of order $\lambda$ or less, and the condition \eqref{traj-cnstr-f} is satisfied.

\subsection{Decoherence function}
The decoherence function Eq.~\eqref{decoherence} appearing in the probabilities \eqref{prob} and
\eqref{prob2} constrains the difference between the central times of detection time as  
\begin{align}
 \delta{\tD}^0 \le 2\bar{\sigma}_{T} \label{decoh-cnstr}.
\end{align}
Here, $\bar{\sigma}_{T_D}$ is the value of Eq.~\eqref{sigTd} at $m_i=0$.
This constraint gives the ``decoherence time'' as a function of the detector position, particles
velocities and wave packet sizes. Note that in our approach, oscillating particles appear as
intermediate, states and only external particles are observed. Therefore the decoherence condition is
given for the detection times of the scattering particle, not for the flight distance of intermediate particles.
From \eqref{decoh-cnstr}, when $\delta {\tD}^0$ is larger than $2\bar{\sigma}_{T_D}$, coherence
is lost and no oscillation is seen.

\subsection{Phase function}
The phase function Eq.~\eqref{phase-func} in the oscillation probability gives a constraint on
oscillation period, $T_D^{osc}$ :
\begin{align}
2\big(2\pi-\Theta_{21}(0)\big)\tilde{\sigma}_{T_D} \le T_D^{osc} \label{phase-cnstr}.
\end{align}
This relation implies that when the oscillation period is smaller than the width of the
detection time, the oscillation disappears.

\subsection{Lifetime}
The lifetime effect from a source particle $A$ is seen explicitly as $\exp(-\Gamma/2 t^0_i)$ in the
absolute square of the amplitude \eqref{prob2}. From the right-hand side of \eqref{amp-si}, 
this term gives constraint
\begin{align}
\delta t^0_1 \le \tau \label{lifetime-cnstr}.
\end{align}
From this relation, to maintain coherence, the difference in the production time 
for intermediate particles must be smaller than the lifetime of the source particle.

\section{The numerical results of transition probabilities}
In this section, we give the results for the numerical calculations of the
oscillation probabilities. The particle $C$, which corresponds to a muon 
in pion decay, usually is not detected in most experiments and
observations. From this fact, the 
oscillation probabilities that we actually measure are the sums of probabilities 
over $\pCz$. Therefore, we consider the following probabilities instead\footnote{
Actually, $|\tilde{S}|^2$ is constant with $\xC$ in finite macroscopic range. Therefore this value
is equivalent to a probability integrated over coordinate
\begin{align}
\int d\vec{p}^0_C \int d\vec{\xC} |\tilde{S}|^2, \nn
\end{align}
in which the orthogonality of states with different values of $\pC^0$ is satisfied.} 
of those in Eq.~\eqref{prob} and Eq.~\eqref{prob2} :
\begin{align}
\bar{P}(X_B)=N_{\text{norm}}\int\!\!d\vec{p}^0_C |\tilde{S}|^2. \label{prob3}
\end{align}
Here, $|\tilde{S}|^2$ is given in Eq.~\eqref{prob} or \eqref{prob2}, and $N_{\text{norm}}$ is a
normalization factor. 
We investigate the following three situations of different parameters:
\begin{enumerate}
 \item Intermediate particles are produced by the decay of the particle $A$ in
       flight. The average momentum of each external particle is
       taken so that the average momentum of the intermediate particle
       is about $430$ MeV. This case mimics long base line 
       experiments. 
 \item The intermediate particles are produced by the decay of the particle $A$ at
       rest. The mass of the particle $A$ is chosen as about $140$ MeV, which
       gives the intermediate particles' momentum as $30$ MeV.
 \item The intermediate particles are produced by the decay of the heavy particle $A$ in
       flight. The average momenta of external particles are taken so that
       the intermediate particles momentum becomes $0.4$ MeV.
\end{enumerate}
The first and the second cases correspond to ``decay in flight'' (DIF) and 
``decay at rest'' (DAR) neutrinos in neutrino oscillation experiments using pion sources. 
The third case correspond to ``solar neutrinos from $^7$Be decay'' whose energies are
MeV (low energy or LE). 

In baseline experiments, the momenta of source particles are focused in the direction 
of the detector and the momenta of produced neutrinos and accompanying charged leptons are 
in almost the same direction. In solar neutrino observations, the velocities of 
source and accompanying particles are slow, because their momenta are much lower than their masses.
Therefore the one-dimensional approximation is valid in the above three cases. For these reasons, we
perform $\pCz$ integral in one dimension.

For the numerical calculations, we use the following parameters:  
$m_1=0.1$ eV and $m_2=0.2$ eV and $\theta=\frac{\pi}{4}$ in
all cases. These are the same values in (I). The wave packet sizes, 
$\sigma_L$ $(L=A, B, C, D)$, are also taken to be the same. 
The values of other parameters are given in Table \ref{param_set}. 
\begin{table}[htb]
\begin{center}
\begin{tabular}{|c||c|c|c|}\hline
 &case 1 & case 2 & case 3\\ \hline\hline 
$m_A$ & $140.0$             & $140.0$              & $6.3\times 10^3$\\ \hline
$m_B$ & $2.9\times 10^{-4}$ & $4.2\times 10^{-3}$ & $2.4\times 10^{-1}$\\ \hline
$m_C$ & $106.0$             & $106.0$              & $6.29960\times 10^3$\\ \hline
$m_D$ & $0.5$               & $0.5$                & $0.5$\\ \hline\hline
$p_A^0$ & $1000.0$            & $0.0$                & $1.3$\\ \hline
$p_B^0$ & $0.0$               & $0.0$                & $0.0$\\ \hline
$p_D^0$ & $428.8$             & $29.9$               & $0.4$\\ \hline\hline
$X_A$ & $0.0$               & $0.0$                & $0.0$ \\ \hline 
$X_C$ & $300.0$             & $-5.0$               & --- \\ \hline
$X_D$ & $X_B+1.0\times 10^{-9}$& $X_B+1.0\times 10^{-9}$& $X_B+1.0\times 10^{-9}$ \\ \hline
$T_C$ & ---                 & ---                  & $\tau$ \\ \hline\hline
$\tau$ & $2.6\times 10^{-8}$ & $2.6\times 10^{-8}$  & $1.0\times 10^{-12}$\\ \hline
\end{tabular}
\caption{The masses $m_L$ and the momenta $p^0_L$ in MeV, and the positions $X_L$ in meters. 
The lifetimes or relaxation times, $\tau$ are in seconds. ($L=A,B,C,D$)}
\label{param_set}
\end{center}
\end{table}

$T_C$ in case 1 and case 2 and $X_C$ in case 3 are not shown in Table.\ref{param_set}.
We set these parameters by hand as a function of $p_C^0$, because there is no way to 
determine both $X_C$ and $T_C$ in one dimension. The concrete forms of $T_C(p^0_C)$ and $X_C(p^0_C)$
are given in the following subsections.

\subsection{Case 1: Decay in flight}
In the first case, the particle $I$ is produced by the decay of the particle $A$ 
in flight accompanying $C$, and the particle $D$ appears
through the interactions between the particle $B$ at rest and $I$. 

In baseline experiments, the particle $C$ is considered to be stopped at a beam
dump. Therefore we set $X_C$ to a constant value. But the detection time of $C$ is unknown. Therefore
we study the probability of a certain time $T_C$, which is a function of $P^0_C$. 
$T_C$ should be such that the time order for the central times is given by
\begin{align}
0&<t^0_{1i}({\tD^0}_i)<t^0_{2i}({\tD^0}_i)<{\tD^0}_i, \label{time_order1} \\
t^0_{1i} &< T_C,\label{time_order2}
\end{align}
where $t^0_{1i}$, $t^0_{2i}$ and ${\tD^0}_i$ are functions of $T_C$.
One choice satisfying this condition is
\begin{align}
T_C(p_C^0)=\frac{X_C}{v_C(p_C^0)}\times 0.999 .\label{tc_dif}
\end{align}
Using Eqs.~\eqref{tc_dif} and \eqref{prob3}, the oscillation probabilities
are calculated numerically. 

\begin{figure}[h]
  \centerline{\includegraphics[width=10cm]{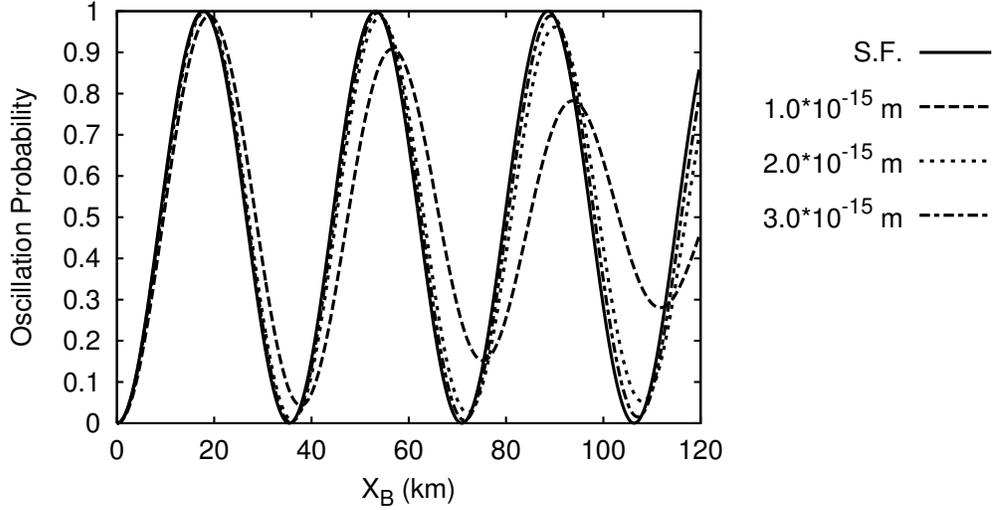}}
  \caption{The DIF oscillation probability with infinite lifetime. The solid
  curve represents the standard formula (S.F.), and the dashed, dotted and  
  dashed-dotted curves correspond to the cases in which the wave packet sizes are  $1.0\times 10^{-15}$
  m, $2.0\times 10^{-15}$ m and $3.0\times 10^{-15}$ m, respectively.The
  horizontal axis is the position of $B$, $X_B$ (km).}
  \label{dif}
\end{figure}

\begin{figure}[h]
  \centerline{\includegraphics[width=10cm]{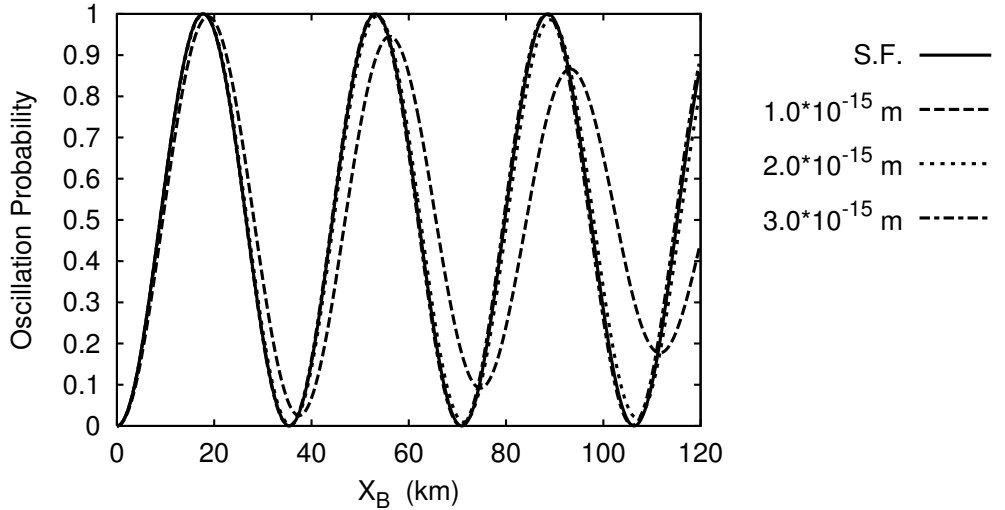}}
  \caption{The DIF oscillation probability with finite lifetime. The solid
  curve represents the standard formula (S.F.), and the dashed, dotted and dashed-dotted
  curves correspond to the cases the wave packet sizes are  $1.0\times 10^{-15}$
  m, $2.0\times 10^{-15}$ m and $3.0\times 10^{-15}$ m, respectively. The
  horizontal axis is the position of $B$, $X_B$ (km).}
  \label{dif_life}
\end{figure}

The oscillation probabilities with an infinite lifetime and a finite lifetime are shown 
in Figs.\ref{dif} and \ref{dif_life}, respectively.
In Figs.~\ref{dif} and \ref{dif_life}, the oscillation length becomes longer than
that of the standard formula when the wave packet sizes are smaller than 
$3.0\times 10^{-15}$ m, and the amplitude of the oscillation probability becomes smaller than $1$ as
$X_B$ becomes larger or as the wave packet sizes become smaller.

\begin{figure}[h]
   \begin{center}
  \begin{tabular}{cc}
   \includegraphics[height=5.5cm]{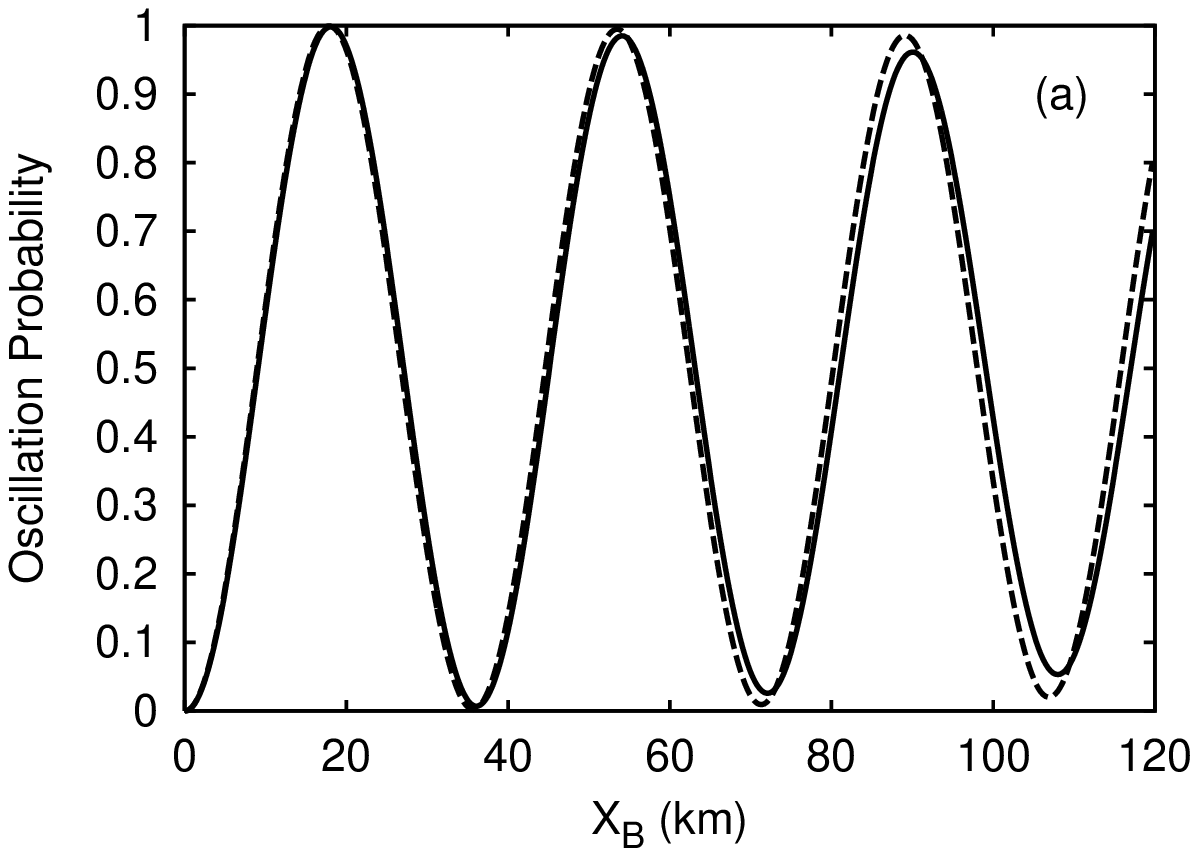}&
   \includegraphics[height=5.5cm]{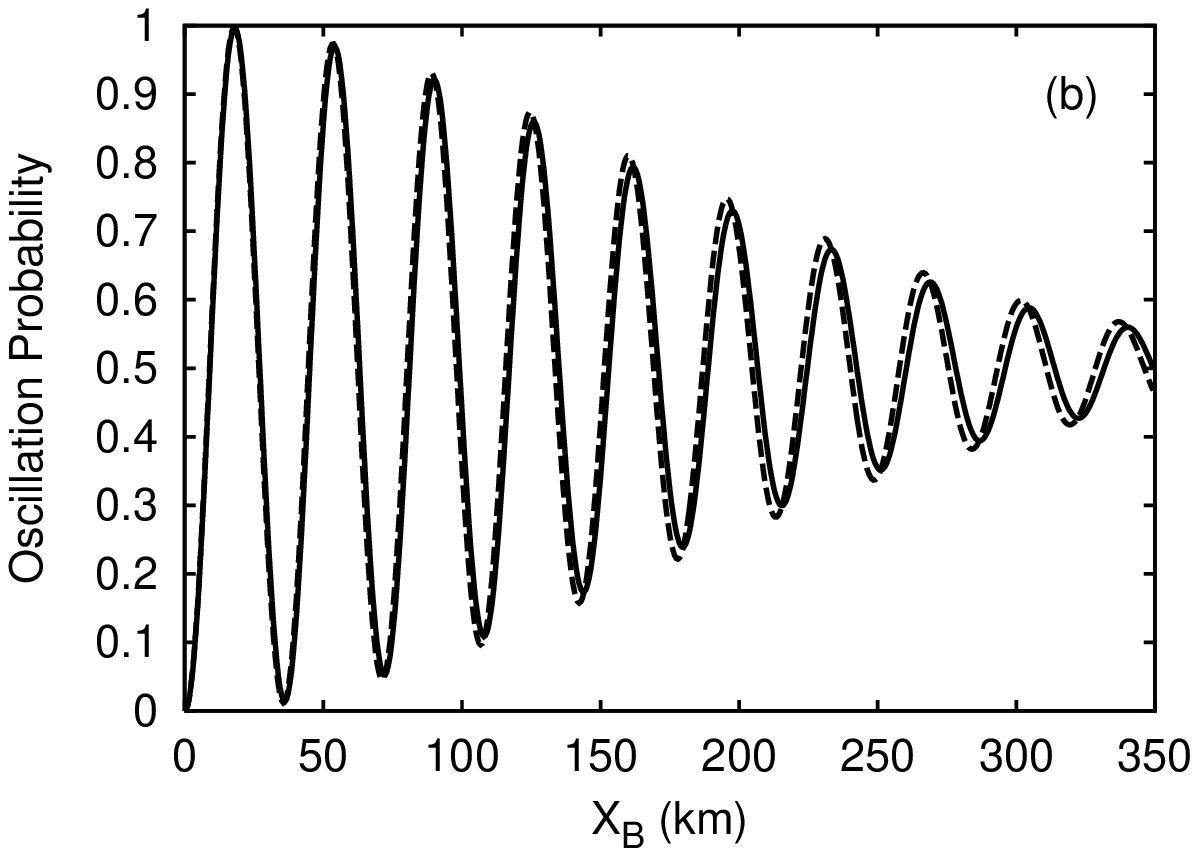} \\
   \includegraphics[height=5.5cm]{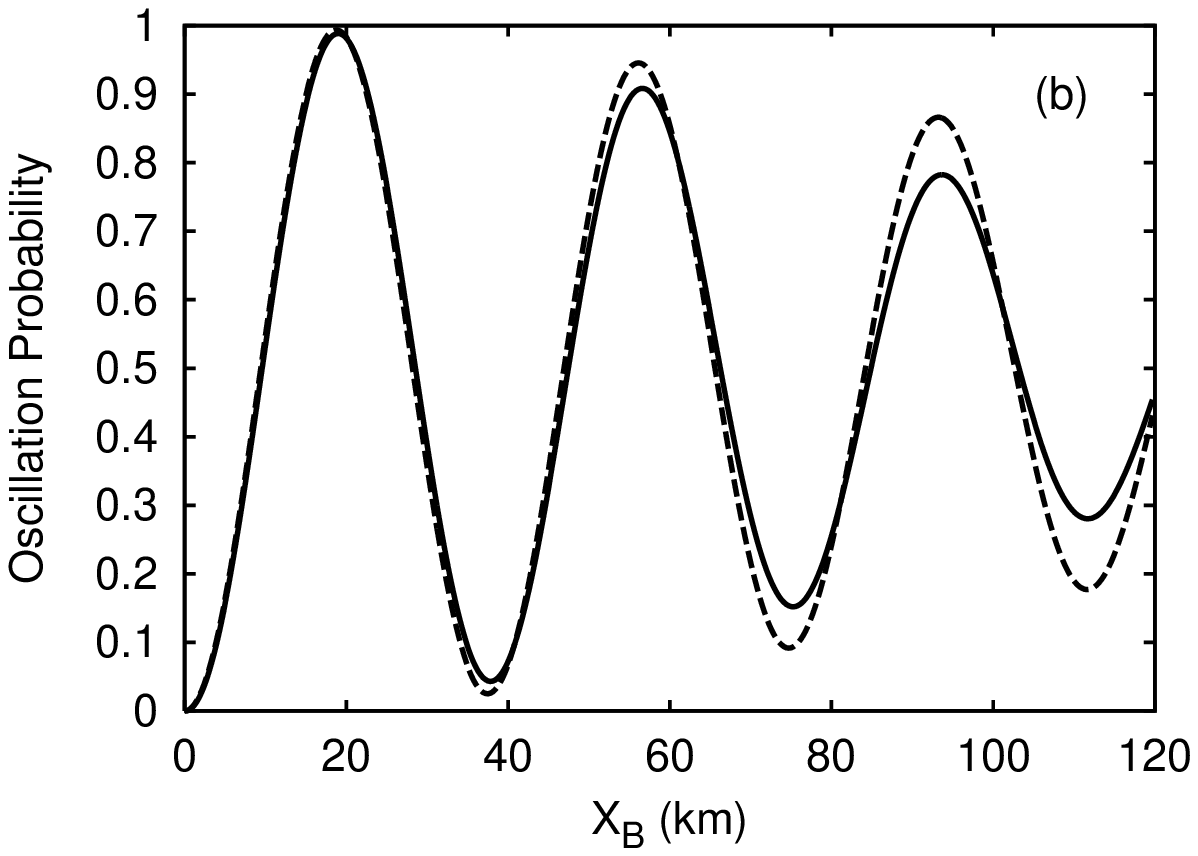}&
   \includegraphics[height=5.5cm]{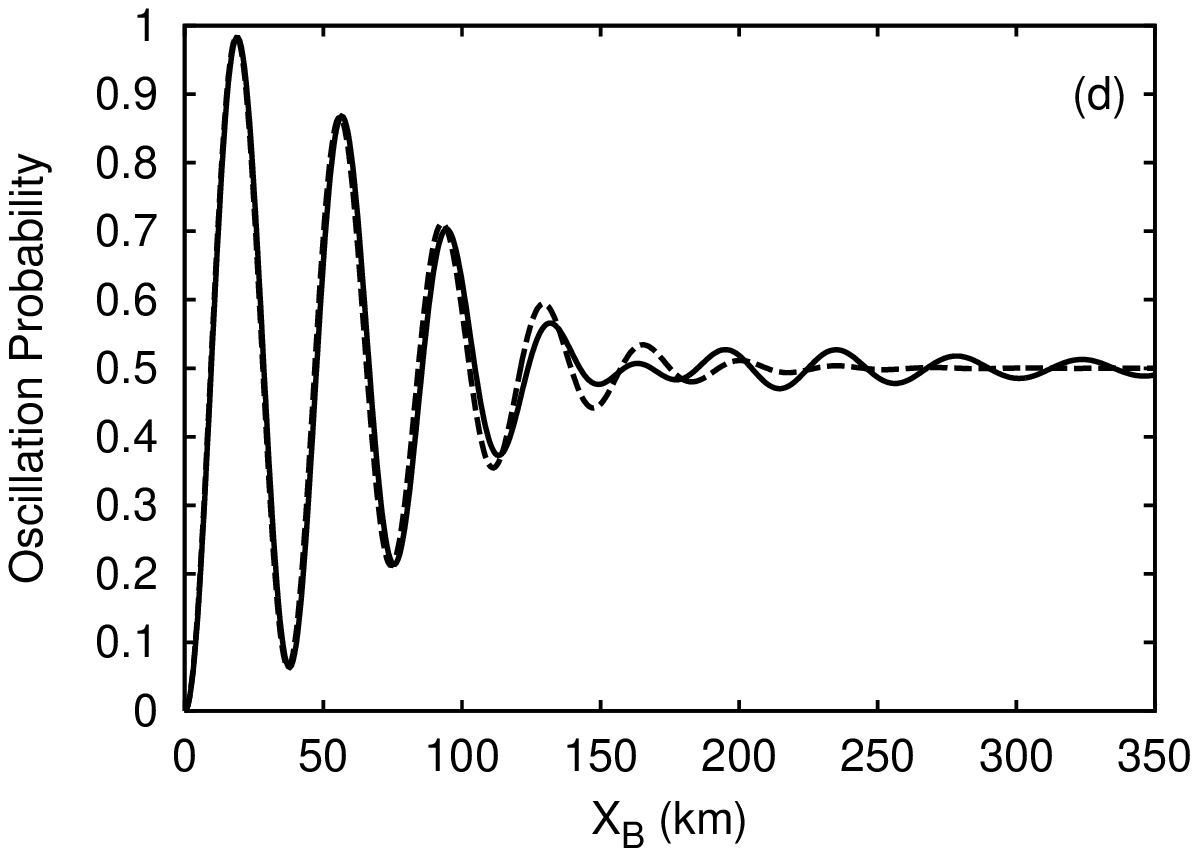} 
  \end{tabular}
 \end{center}
 \caption{The oscillation probabilities at $\sigma_x=2.0\times 10^{-15}$ m in (a) and (b), 
 and those at $\sigma_x=1.0\times 10^{-15}$ m in (c) and (d). 
 The solid curves represent the oscillation probabilities with infinite lifetime and the dashed curves
 represent those with finite lifetime.}
 \label{dif_prob-comp}
\end{figure}

In Figs.~\ref{dif_prob-comp} (a), (b), (c) and (d), we compare the oscillation probabilities
of the infinite lifetime with those of the finite lifetime at
$\sigma_x=2.0\times 10^{-15}$ m for (a) and (b) and $\sigma_x=1.0\times 10^{-15}$ m for (c) and (d).
From Fig.~\ref{dif_prob-comp}, it is seen that the oscillation length is longer 
and the amplitude of oscillation is smaller when the lifetime is infinite.

The increase of oscillation length is caused by the increase of the time widths, 
$\bar{\sigma}_{t_{1i}}$ and $\bar{\sigma}_{Ti}$, in the coefficients in
Eq.~\eqref{prob}.
In the following, we clarify the reason that the oscillation length increases with the time widths. 

\begin{figure}[h]
 \begin{center}
  \begin{tabular}{cc}
   \includegraphics[height=5.5cm]{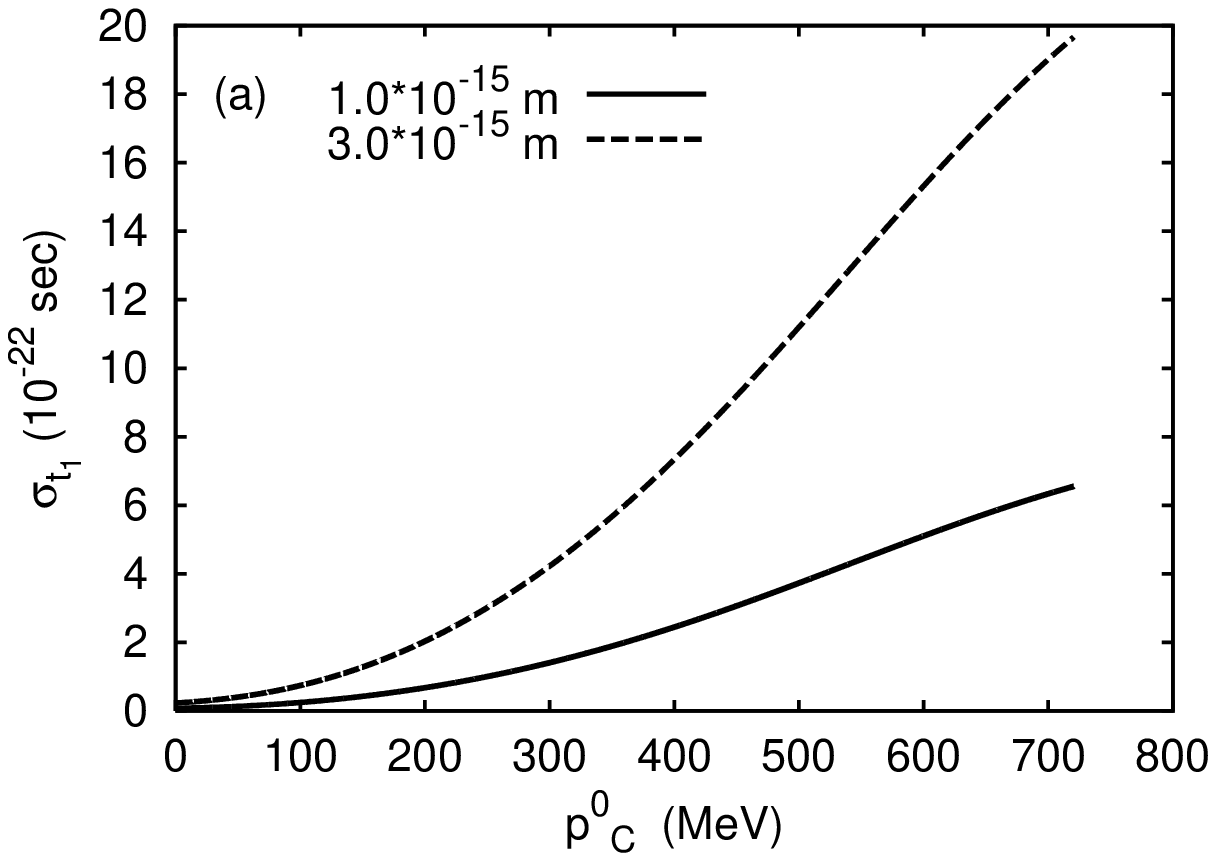}&
   \includegraphics[height=5.5cm]{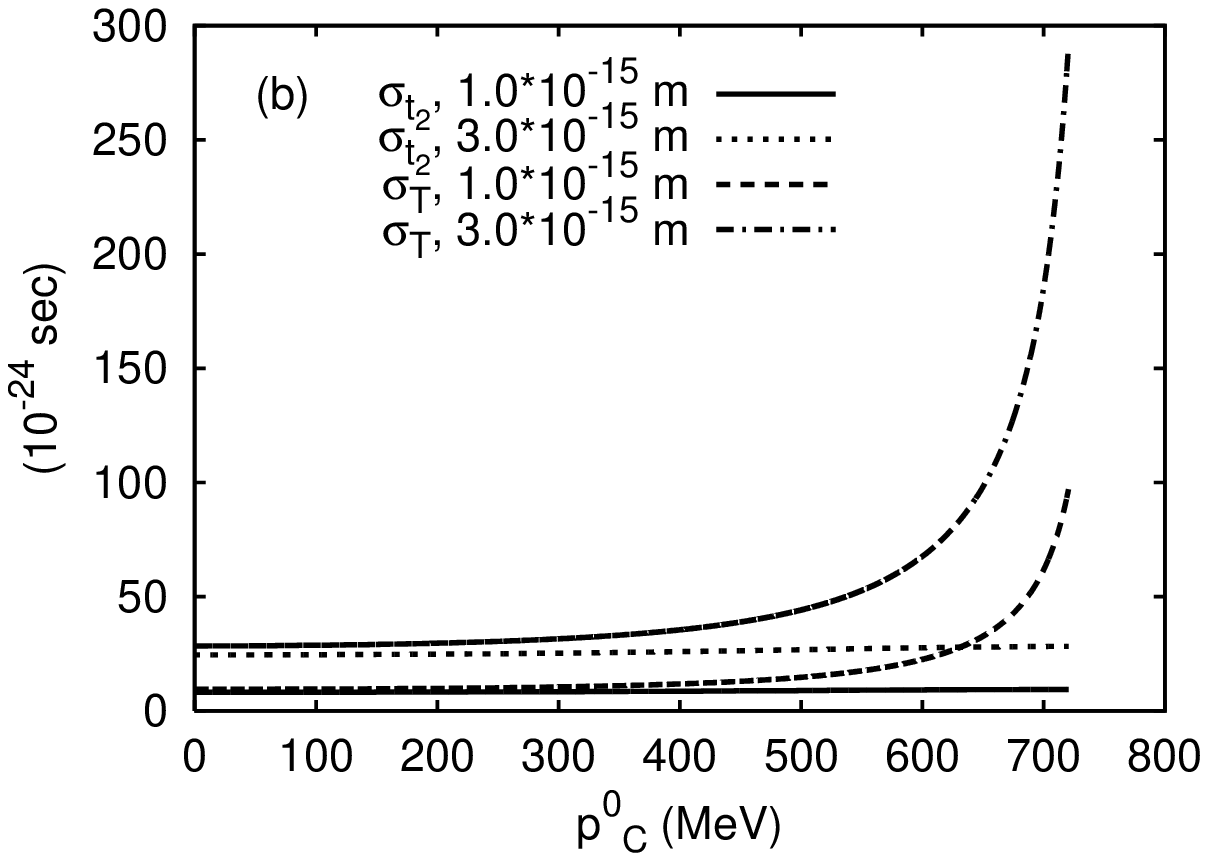}
  \end{tabular}
 \end{center}
 \caption{(a): The $p^0_C$ dependence of $\bar{\sigma}_{t_1}$. The solid curve corresponds to 
 $\bar{\sigma}_{t_1}$ at $\sigma_x=1.0\times 10^{-15}$ m, and the dashed curve corresponds to 
 $\sigma_x=3.0\times 10^{-15}$ m. (b) :The $p^0_C$ dependence of $\bar{\sigma}_{t_2}$ and $\bar{\sigma}_T$. 
 The solid and dotted curves represent $\bar{\sigma}_{t_2}$ at $\sigma_x=1.0\times 10^{-15}$ m
 and $\sigma_x=3.0\times 10^{-15}$ m, and the dashed and dashed-dotted curves are $\bar{\sigma}_T$ at
 $\sigma_x=1.0\times 10^{-15}$ m and $\sigma_x=3.0\times 10^{-15}$ m.}
 \label{dif_time_widths}
\end{figure}

In Fig.~\ref{dif_time_widths}, the $p^0_C$ dependences of the time widths are shown.
From Figs.~\ref{dif_time_widths} (a) and (b), it is seen that
$\bar{\sigma}_{t_{1i}}$ and $\bar{\sigma}_{T_i}$ grow with $p^0_C$. 
From Eq.~\eqref{trj_h}, it is seen that $H$ loses its $t_1$ dependence when $v_C$ equals 
$v_A$. As we mentioned in the last part of Section 4, this growth of the time widths shifts the peak
of the oscillation probability upward from that of the Gaussian in Eq.~\eqref{prob}.

\begin{figure}[h]
 \begin{center}
  \begin{tabular}{cc}
   \includegraphics[height=5.5cm]{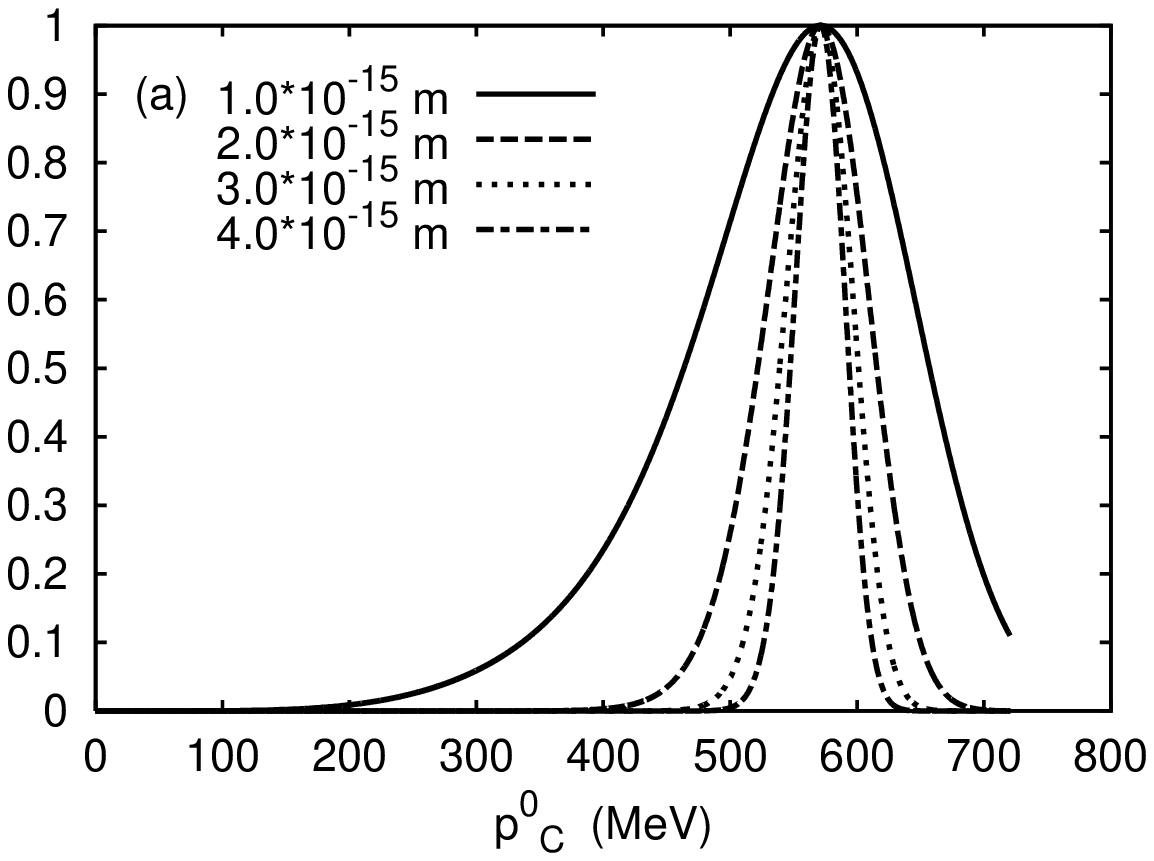}&
   \includegraphics[height=5.5cm]{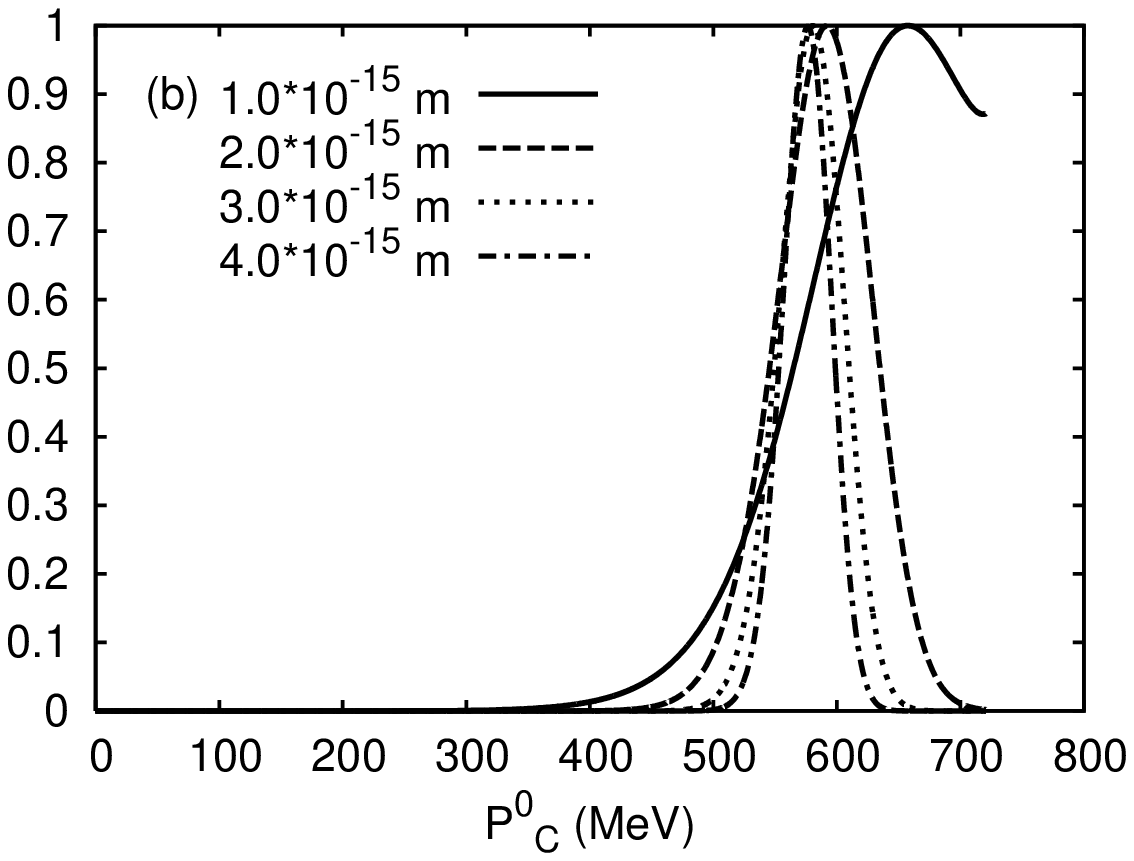}
  \end{tabular}
 \end{center}
 \caption{(a): The $p^0_C$ dependence of the Gaussian function. 
 (b): The $p^0_C$ dependence of the first term in Eq.~\eqref{prob}. 
 The solid curve represents $\sigma=1.0\times 10^{-15}$ m, 
 and the dashed, dotted and dashed-dotted curves represent $2.0\times 10^{-15}$,
 $3.0\times 10^{-15}$ and $4.0\times 10^{-15}$ m, respectively, in both graphs.}
 \label{dif_peak}
\end{figure}

Figure \ref{dif_peak} (a) shows the Gaussian function $\exp(2A_i)$ in the infinite lifetime case, 
and Fig.~\ref{dif_peak} (b) shows the absolute square of total amplitude Eq.~\eqref{prob}, in which the
maximum values are normalized to unity, at 
$\sigma=1.0$, $2.0$, $3.0$, $4.0\times 10^{-15}$ m, respectively.
From Fig.~\ref{dif_peak} (a), it is seen that the peaks of the Gaussian function are  
independent of the wave packet sizes. From Fig.~\eqref{dif_peak} (b), it is seen that the 
peaks of the oscillation probability depend on the wave packet sizes. For
large $\sigma_x$ ($> 4.0 \times 10^{-15}$ m), the peaks of the oscillation probability are the 
same. But for small $\sigma_x$ ($< 4.0 \times 10^{-15}$ m), the peaks of the oscillation probability 
are different from those of the Gaussian functions. This happens because the time widths grow with
$p^0_C$ in Fig.~\ref{dif_time_widths} and the width of the Gaussian function becomes large in
Fig.~\ref{dif_peak} (a). Then, the main contribution to the oscillation phase, Eq.~\eqref{phase_dif} 
in Eq.~\eqref{prob3}, comes from the higher $p_C^0$ region.

\begin{figure}[t]
  \centerline{\includegraphics[width=9cm]{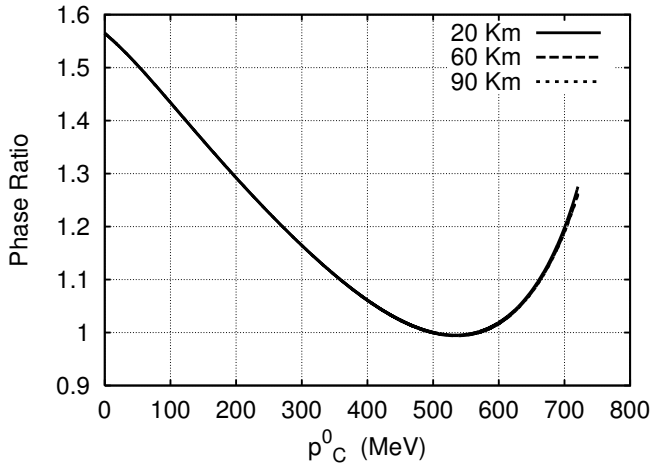}}
  \caption{The $p_C^0$ dependence of the ratio of oscillation phases. 
 The solid, dashed and dotted curves represent the ratios at $X_B=20$, $60$, $90$
 km, respectively.}
  \label{dif_phase-ratio}
\end{figure}

Figure \ref{dif_phase-ratio} displays the $p_C^0$ dependence of the ratio of the phase
difference Eq.~\eqref{phase_dif} to that of the standard formula for $X_B=20$, $60$ and $90$ km : 
\begin{align}
\text{Phase Ratio}=\Theta_{21}(\tilde{T}^0_D)/\left(\frac{\Delta m^2_{21}}{2 E}X_B\right).
\label{phase-ratio}
\end{align}
Here $X_B$ is regarded as being the same as the travel distance of intermediate particles, 
and the energy $E$ in the phase of the standard formula is determined by $\Delta P=0$ and 
$\Delta \tilde{E}^0_1 \simeq \Delta \tilde{E}^0_2\simeq 0$.
From Fig.~\ref{dif_phase-ratio}, it is seen that the ratio is almost independent
of $X_B$ but depends on $p_C^0$. As is shown in Eq.~\eqref{phase_std}, the phase ratio should become
$1$ at the solution $\Delta P=0$ and $\Delta \tilde{E}^0_1=\Delta \tilde{E}^0_2=0$, 
$p^0_C=570$ MeV. When $\sigma_x$ is larger than $3.0\times 10^{-15}$ m, the main contribution 
to the integral in Eq.~\eqref{prob3} comes from the region where the energy and momentum are conserved.
Equation \eqref{prob3} is almost the same as the standard formula.
But from Fig.~\ref{dif_peak}, when $\sigma_x$ is smaller than $3.0\times 10^{-15}$ m, the 
peak of the Gaussian part in the integrand moves to higher $p^0_C$ region than that of 
Fig.~\ref{dif_peak} (a), and the main contribution to 
Eq.~\eqref{prob3} comes from the higher $p^0_C$ region. Then, the oscillation length
becomes longer than that of the standard formula.

In the case that the particle $A$ has a finite lifetime, the main contribution to
the integral comes from the lower $p^0_C$ region, because of the presence of the lifetime in 
Eq.~\eqref{amp-si}. 

\begin{figure}[h]
 \begin{center}
  \begin{tabular}{cc}
   \includegraphics[height=5.5cm]{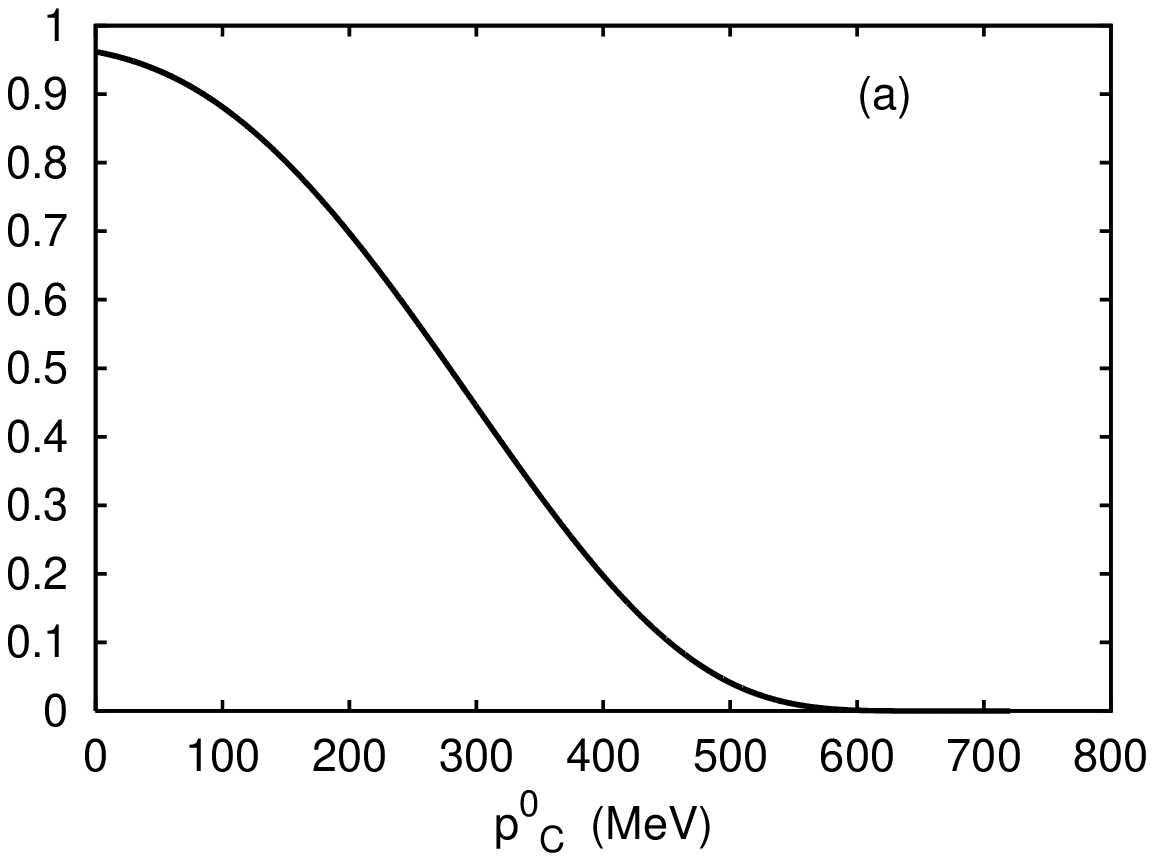}&
   \includegraphics[height=5.5cm]{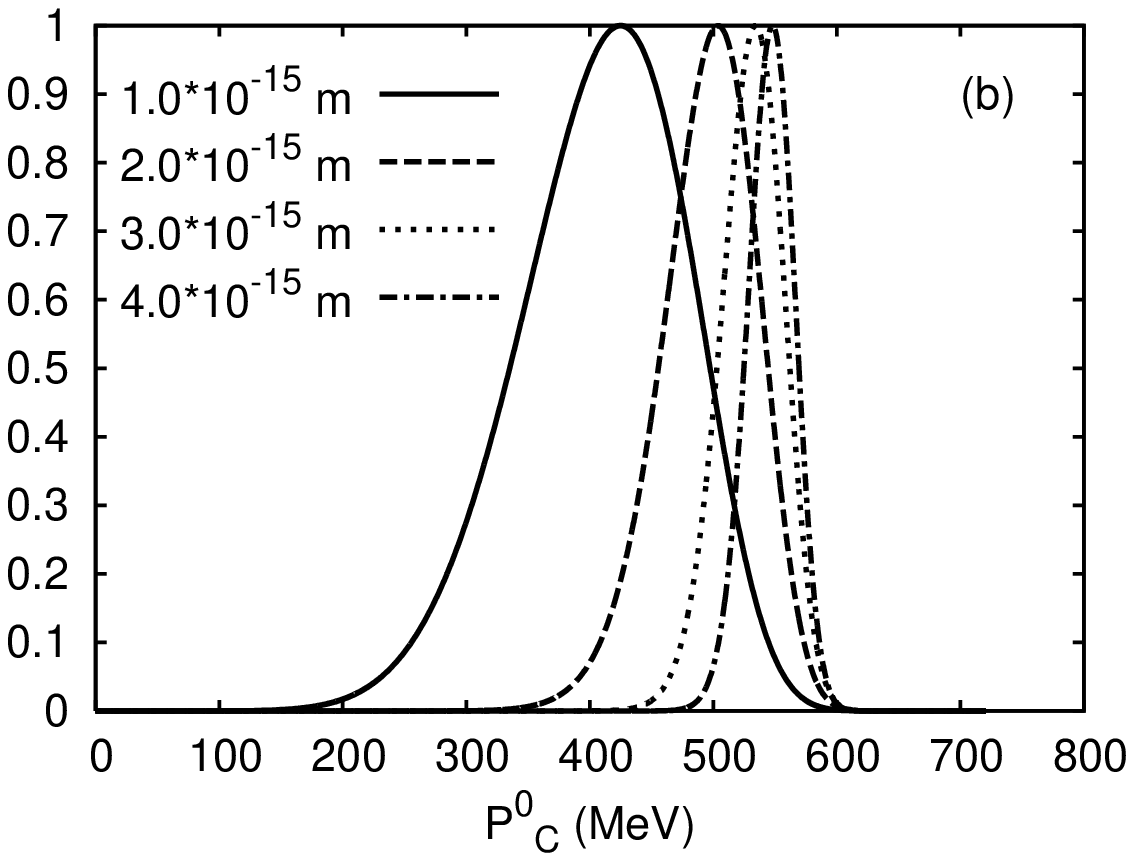}
  \end{tabular}
 \end{center}
 \caption{(a): $\exp(-\Gamma t^0_{1i})$ in the exponent of
 Eq.~\eqref{prob2}. (b): The first term in Eq.~\eqref{prob2}. The solid and 
 dashed curves correspond to $1.0\times 10^{-15}$ m and
 $2.0\times 10^{-15}$ m, and the dotted and dashed-dotted
 curves correspond to $3.0\times 10^{-15}$ m and $4.0\times 10^{-15}$ m, respectively.}
 \label{dif_life_peak}
\end{figure}

Figure \ref{dif_life_peak} (a) displays the $p^0_C$ dependence of 
$\exp(-\Gamma t^0_{1_i})$, and (b) displays the $p^0_C$ dependence of the absolute square of the
amplitude \eqref{prob2}, in which the maximum values are normalized to unity, 
at $\sigma=1.0,\ 2.0,\ 3.0,\ 4.0\times 10^{-15}$ m. The quantity $\exp(-\Gamma t^0_{1_i})$ is almost independent of
$X_B$ and $\sigma_x$. From Fig.~\ref{dif_life_peak} (b), the peaks of the oscillation probability Eq.~\eqref{prob2}
shift to the lower $p^0_C$ region as the wave packet sizes become small.

The phase ratio Eq.~\eqref{phase-ratio} with finite lifetime is almost
the same as that with infinite lifetime, because the lifetime $\tau$ is much larger
than the time widths. Therefore, from Fig.~\ref{dif_phase-ratio}, the oscillation
length is longer than that of the standard formula when the wave packet sizes are
smaller than $2.0\times 10^{-15}$ m  and are slightly smaller than that of
the standard formula when $\sigma_x$ is larger than $3.0\times 10^{-15}$ m.

\begin{figure}[h]
 \begin{center}
   \includegraphics[height=7.0cm]{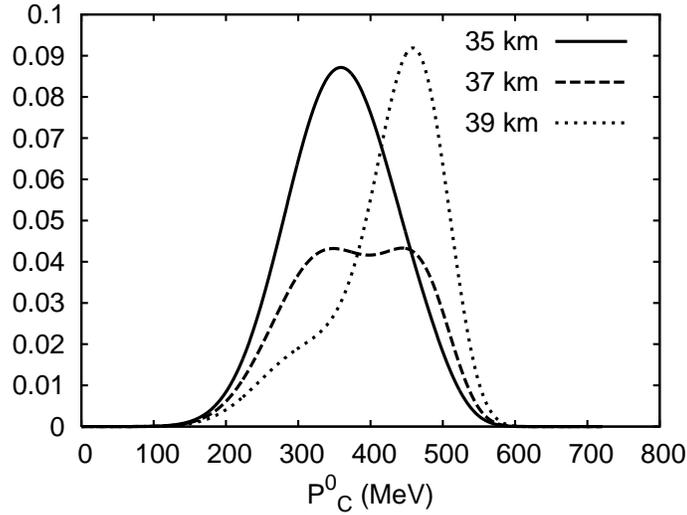}
 \end{center}
 \caption{$p^0_C$ dependence of the absolute square of the flavor changing amplitude at
 $\sigma_x=1.0\times 10^{-15}$ m. The vertical axis is normalized by the value of $|S|^2$ at
 $X_B=18$ km.
 The solid curve represents the flavor conserving process at $X_B=35$ km, and the dashed and dotted  
 curves represent the flavor changing processes at $X_B=37$, $39$ km.}
 \label{dif_life_s-change}
\end{figure}

Figure \ref{dif_life_s-change} displays the $p^0_C$ dependence of the absolute square of the flavor
changing amplitudes around $X_B=37$ km. It is seen that the peaks of the absolute square of
the amplitudes change and the shape of each amplitude is deformed with distance. 
The Gaussian shapes of the amplitudes are deformed by the cosine in the oscillation term. 
The behavior of the oscillation depends on $X_B$ as well as $p^0_C$, since the phase is 
proportional to $X_B/|\vec{k}^0|$. Then, as a result, the change and deformation given in
Fig.~\ref{dif_life_s-change} occur. This effect is seen clearly around the oscillation minimum.

\subsection{Case 2 : Decay at rest}
In the DAR, the intermediate particles $I$ and particle $C$ are 
produced by the decay of particle $A$ at rest. The particle $I$ 
is scattered by the particle $B$ at rest, and then the particle $D$ appears.

For the same reason as in case 1, $X_C$ is set to  $-5.0$ m and $T_C$ is given as
\begin{align}
T_C(p^0_C)=\frac{X_C}{v_C(p^0_C)}\times 2.0 \label{tc_dar}.
\end{align}
Because the source particle $A$ is at rest, the numerical factor in $T_C(p^0_C)$ must be greater than
$1$ to satisfy the conditions \eqref{time_order1} and \eqref{time_order2}. Here we take this factor
to be $2$.

\begin{figure}[ht]
 \centerline{\includegraphics[width=10.0cm]{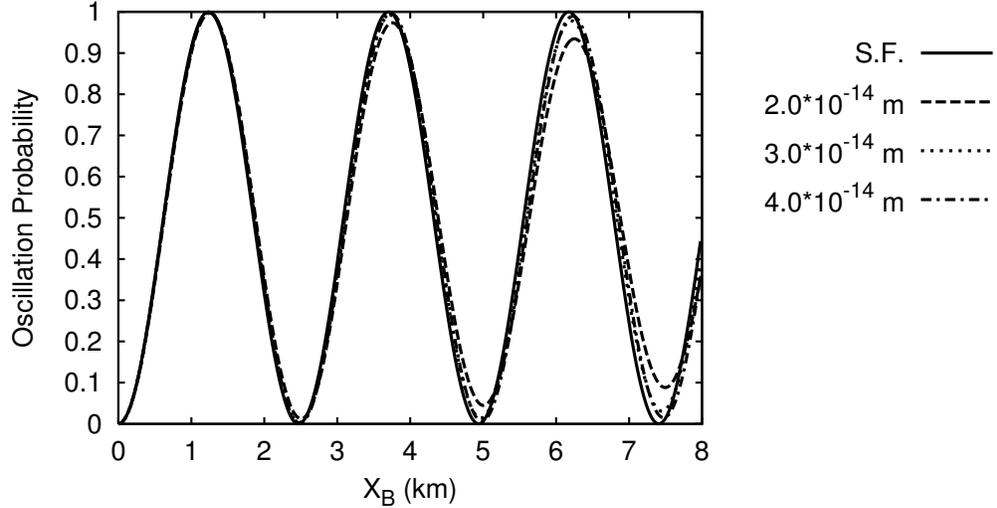}}
  \caption{The DAR oscillation probability with infinite lifetime: The solid
  curve represents the standard formula (S.F.), and the dashed, dotted and dashed-dotted 
  curves correspond to the cases in which the wave packet sizes are  $2.0\times 10^{-14}$
  m, $3.0\times 10^{-14}$ m and $4.0\times 10^{-14}$ m, respectively. The
  horizontal axis is the position of $B$, $X_B$ (km).}
 \label{dar}
\end{figure}

The oscillation probabilities with finite lifetime are shown in Fig.~\ref{dar}.
It is seen again that the amplitude of the oscillation becomes smaller and the period of
the oscillation probability becomes longer than that of the standard formula as the wave packet
sizes become smaller than $4.0\times 10^{-14}$ m.

In this case, $t^0_{1i}$ is almost zero, since the particle $A$ is at rest. 
Because of this, there are no significant differences between the source particle with a finite 
lifetime and an infinite lifetime.

\subsection{Case 3 : Low energy}
In the last case, the intermediate particles are produced by the decay of the
heavy particle $A$ in flight. The central value of its momentum is about $1.3$ MeV,  which is much lower
than in case 1. The particles $A$ and $C$ have larger masses and 
smaller momenta than in the other two cases.

In contrast to the above two cases, this case corresponds to the solar neutrinos from $^7$Be decays. 
The change of quantum mechanical states by scatterings is considered to be equivalent to 
detection or observation. Therefore the detection time of $C$
is taken to the relaxation time, which is assumed to be the same value of particle $A$, and 
$X_C$ is given as a function of $p^0_C$. The quantities $T_C$ and $X_C$ are
\begin{align}
T_C&=\tau, \\
X_C(p^0_C)&=1.5\times T_C\ v_C(p^0_C),
\end{align}
where $\tau$ is the relaxation time of the particle $A$, and its value is set to
$10^{-12}$ sec.

\begin{figure}[ht]
 \centerline{\includegraphics[width=10.0cm]{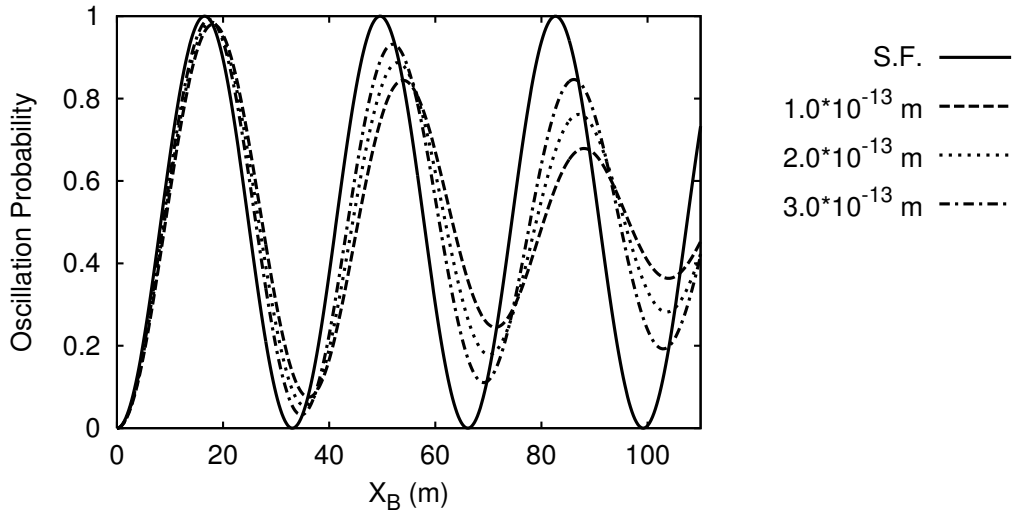}}
  \caption{The LOW oscillation probability with lifetime: The solid
  curve represents the standard formula (S.F.), and the dashed, dotted and dashed-dotted 
  curves correspond to the case in which the wave packet sizes are  $1.0\times 10^{-13}$
  m, $2.0\times 10^{-13}$ m and $3.0\times 10^{-13}$ m, respectively. The
  horizontal axis is the position of $B$, $X_B$ (m).}
 \label{low}
\end{figure}

In Fig.~\ref{low}, the oscillation probability with finite lifetime is shown.
In this case, the oscillation length becomes longer and the oscillation amplitude becomes
smaller again. The wave packet effect becomes observable if the wave packet sizes are 
of order $10^{-13}$ m. A wave packet size of order $10^{-13}$ m is larger than those for the DIF and
DAR cases. This is because the central values of the momenta are
lower than in the DIF and DAR cases.

\section{Discussion and conclusion} 
In this paper, we have studied particle oscillations in  the
intermediate state of transition amplitudes of wave packets based on a
simple scalar model. A source particle, a target particle, and two 
scattered particles are the wave packets. In this situation, wave 
functions do not spread infinitely, but are localized within a finite 
width $\sigma_x$, and interference occurs only among 
intermediate particles that overlap spatially. This interference
disappears and the oscillation probability deviates from those of the standard
formula in parameter regions where the intermediate particles are separated spatially. 

We computed the total oscillation probability and the phase factor 
of the amplitude numerically. We found that the oscillation probability agrees with that of the
standard formula if the wave packet sizes are of semi-macroscopic values, and that the oscillation
probability  deviates from those of the standard 
formula in extreme parameter regions. This occurs when the wave packet
sizes are of the  order of $10^{-13}$ m  or smaller. In this
region, the wave packet size effects become visible. The oscillation
amplitude becomes smaller and the oscillation period becomes larger  than
those of the standard formula. We hope that this region may  be realized
experimentally and that the modified formula found here will be tested in the 
future. 

Although our results were obtained on the basis of a simple scalar model, we 
hope that for  precision measurements of the neutrino parameters
such as masses, the MNS matrix, and others, our considerations will be 
valuable.

\section*{Acknowledgements}
This work was partially supported by a special Grant-in-Aid for the Promotion of Education and
Science from Hokkaido University and a Grant-in-Aid for Scientific Research on Priority Areas 
(Dynamics of Superstrings and Field Theories, Grant No.13135201), and a Grand-in-Aid for Scientific 
Research on Priority Areas (Progress in Elementary Particle Physics of the 21st Century through
Discoveries of Higgs Boson and Supersymmetry, Grand No. 16081201 ), provided by the Ministry of
Education, Culture, Sports, Science and Technology, Japan, and the Nukazawa Science Foundation.

\appendix
\section{The Center Times and the Time Widths}\label{A}
In this appendix, we give the explicit forms of the time widths 
($\sgTf, \sgTs, \sgT{i}$) and the center times ($t^0_{1i}, t^0_{2i}, {\tD^0}_i$). 
We omit the mass index $i$ for simplicity. However, we note that one can easily
obtain the center times and the time widths for specific mass eigenstates 
by replacing $\nu_I$ with $\nu_i$.

The time widths and the center times are written 
in terms of the time derivatives of the classical trajectory.
The classical trajectory given in Eq.~\eqref{cls_trj} is
\begin{align}
Z_i(t_1,t_2,\tD)&=\frac{\sg_{AC}\sg_{BD}}{\sg}\vec{F}^2_i(t_1,t_2,\tD)
 +\frac{\sg_B\sg_D}{\sg_{BD}}\vec{G}^2(t_2,\tD)+\frac{\sg_A\sg_C}{\sg_{AC}}\vec{H}^2(t_1), \\
\vec{F}_i(t_1,t_2,\tD)&=\vec{x}^0_2(t_2,\tD)-\vec{x}^0_1(t_1)-\vec{v}_i(t_2-t_1), \label{traj-f} \\
\vec{G}(t_2,\tD)&=\xD-\xB-t_2\vB-(\tD-t_2)\vD, \label{traj-g} \\
\vec{H}(t_1)&=\xC-\xA-t_1\vA-(\tC-t_1)\vC.\label{traj-h}
\end{align}
We use following abbreviated expressions for the time
derivatives of the classical trajectory :
\begin{align}
Z_{11} &\equiv \sdel{Z(t_1,t_2,\tD)}{t_1} =2\frac{\sg_{AC}\sg_{BD}}{\sg}\vec{F}^2_1+2\frac{\sg_A\sg_C}{\sg_{AC}}\vec{H}^2_1,\\
Z_{22} &\equiv \sdel{Z(t_1,t_2,\tD)}{t_2} =2\frac{\sg_{AC}\sg_{BD}}{\sg}\vec{F}^2_2+2\frac{\sg_B\sg_D}{\sg_{BD}}\vec{G}^2_2,\\
Z_{TT} &\equiv \sdel{Z(t_1,t_2,\tD)}{\tD} =2\frac{\sg_{AC}\sg_{BD}}{\sg}\vec{F}^2_T+2\frac{\sg_B\sg_D}{\sg_{BD}}\vec{G}^2_T,\\
Z_{12} &\equiv \ddel{Z(t_1,t_2,\tD)}{t_1}{t_2} =2\frac{\sg_{AC}\sg_{BD}}{\sg}\vec{F}_1\cdot\vec{F}_2,\\
Z_{1T} &\equiv \ddel{Z(t_1,t_2,\tD)}{t_1}{\tD} =2\frac{\sg_{AC}\sg_{BD}}{\sg}\vec{F}_1\cdot\vec{F}_T,\\
Z_{2T} &\equiv \ddel{Z(t_1,t_2,\tD)}{t_2}{\tD}
 =2\frac{\sg_{AC}\sg_{BD}}{\sg}\vec{F}_2\cdot\vec{F}_T+2\frac{\sg_B\sg_D}{\sg_{BD}}\vec{G}_2\cdot\vec{G}_T,
\end{align}
and
\begin{align}
Z_{10} &\equiv \left.\del{Z(t_1,t_2,\tD)}{t_1}\right|_{t_1=0 \atop t_2=0}
 =2\frac{\sg_{AC}\sg_{BD}}{\sg}\vec{F}_1\cdot(\vec{F}_0+\vec{F}_T \tD)+2\frac{\sg_A\sg_C}{\sg_{AC}}\vec{H}_1\cdot\vec{H}_0,\\
Z_{20} &\equiv \left.\del{Z(t_1,t_2,\tD)}{t_2}\right|_{t_1=0 \atop t_2=0}
 =2\frac{\sg_{AC}\sg_{BD}}{\sg}\vec{F}_2\cdot(\vec{F}_0+\vec{F}_T \tD)
 +2\frac{\sg_B\sg_D}{\sg_{BD}}\vec{G}_2\cdot(\vec{G}_0+\vec{G}_T \tD),\\
Z_{T0} &\equiv \left.\del{Z(t_1,t_2,T_D)}{\tD}\right|_{
\begin{subarray}{l}
t_1=t^0_1(0)\nn\\
t_2=t^0_2(0)\nn\\
\tD=0\nn
\end{subarray}
}
 =2\frac{\sg_{AC}\sg_{BD}}{\sg}\vec{F}_T\cdot(\vec{F}_0+\vec{F}_1 t^0_1(0)+\vec{F}_2 t^0_2(0))\nn\\
 &\hspace{4.5cm}+2\frac{\sg_B\sg_D}{\sg_{BD}}\vec{G}_T\cdot(\vec{G}_0+\vec{G}_2 t^0_2(0)).
\end{align}
Here, $\vec{F}_i$, $\vec{G}_i$ and $\vec{H}_i\ (i=0,1,2,T)$ are the
coefficient vectors of $t_1$, $t_2$ and $T$, and are given as
\begin{align}
\vec{F}_0&=\frac{\sg_B\xB+\sg_D\xD}{\sg_{BD}}-\frac{\sg_A \xA +\sg_C(\xC-\tC\vC)}{\sg_{AC}}, \\
\vec{F}_1&=-\frac{\sg_A \vA +\sg_C\vC}{\sg_{AC}}+\vec{v}_I,\quad
\vec{F}_2=\frac{\sg_B\vB+\sg_D\vD}{\sg_{BD}}-\vec{v}_I,\quad
\vec{F}_T=-\frac{\sg_D\vD}{\sg_{BD}}, \\
\vec{G}_0&=\xD-\xB,\quad
\vec{G}_2=-\vB+\vD,\quad
\vec{G}_T=-\vD, \\
\vec{H}_0&=\xC-\xA-\tC\vC, \quad
\vec{H}_1=-\vA+\vC.
\end{align}

Using these functions, the time widths and the center times are
written as follows :
\begin{align}
\frac{1}{\bar{\sigma}^2_{t_1}}&=\frac{1}{2}Z_{11},\\
\frac{1}{\bar{\sigma}^2_{t_2}}&=\frac{1}{2}Z_{22}-\frac{1}{4}\bar{\sigma}^2_{t_1} Z_{12}^2,\\
\Delta t^0_1&=-\frac{1}{2}\bar{\sigma}^2_{t_1} \bar{\sigma}_{t_2}Z_{12},\\
\frac{1}{\bar{\sigma}^2_T}&=\frac{1}{2}Z_{TT}
 -\frac{1}{4}(\bar{\sigma}^2_{t_1}+\Delta t^0_1)Z_{1T}^2
 -\frac{1}{2}\bar{\sigma}_{t_2}\Delta t^0_1 Z_{1T} Z_{2T}
 -\frac{1}{4}\bar{\sigma}^2_{t_2}Z_{2T},
\end{align}
and
\begin{align}
t^0_1(T_D)&=-\frac{1}{2}(\bar{\sigma}^2_{t_1}+{\Delta t^0_1}^2)Z_{10}
 -\frac{1}{2}\bar{\sigma}_{t_2}\Delta t^0_1 Z_{20}, \\
t^0_2(T_D)&=-\frac{1}{2}\bar{\sigma}_{t_2}\Delta t^0_1 Z_{10}
 -\frac{1}{2}\bar{\sigma}^2_{t_2}Z_{20}, \\
\tD^0&=-\frac{1}{2}\bar{\sigma}^2_T Z_{T0}.
\end{align}

\section{Measurement of Transition Probability}\label{B}
Following the standard interpretation of measurement in quantum mechanics, the square of the
absolute value of the amplitude, $|S|^2$, gives the transition probability from the initial
state \eqref{ini_state}, which is prepared at $t=0$, to the final state Eq.~\eqref{fin_state}, which is
defined at $t=T_D$. For the probability interpretation to make sense, each observation at the final
state should be made independently and exclusively. This is, when one value is observed for an
observation, the other value should not be observed. The state of one value is different from the state
of a different value. Conversely, the final state should be different if the corresponding values are
different.

Because the orthogonality of states described by the wave packet is peculiar, the total probability
should be defined in a manner that is consistent with experiments. This problem has been solved in
usual scatterings, where the wave packet effects are negligibly small. The detector used in
experiments has a finite macroscopic size, and the total rate observed in a macroscopic detector is
computed with a continuous momentum. The total observed probability is obtained by integrating the
square of the absolute value of the amplitude with a continuous momentum. Therefore this problem has
been solved in the case of normal scattering.

In neutrino scattering, the energy scale is very small, and the event rate is also very
small. Events occur so infrequently that the detector is in different quantum states that are
orthogonal to each other. Also, the detector has a macroscopic size. Consequently, the total rate observed
in a macroscopic detector within a finite detection time is computed by integrating the probability with
the central value of the momentum and the time.

\end{document}